\documentclass[aps,reprint,amsmath,amssymb,superscriptaddress,showpacs,prb]%
{revtex4-2}
\usepackage{graphicx}
\usepackage{color}
\usepackage{lipsum}
\usepackage{dcolumn}
\usepackage{bm}

\newcommand{\Ai}{\mathop{\rm Ai}}
\newcommand{\Bi}{\mathop{\rm Bi}}

\newcommand{\veca}{\bm{a}}
\newcommand{\vecl}{\bm{l}}
\newcommand{\vecm}{\bm{m}}
\newcommand{\veck}{\bm{k}}

\newcommand{\veczero}{\bm{0}}
\newcommand{\vecalpha}{\bm{\alpha}}

\begin{document}

\title{Quantum states in disordered media. I. Low-pass filter approach }

\author{F.~Gebhard}
\email{florian.gebhard@physik.uni-marburg.de}
\author{A.~V.~Nenashev}
\thanks{On leave of absence from Rzhanov Institute of Semiconductor Physics
  and the Novosibirsk State University, Russia}
\affiliation{Department of Physics and Material Sciences Center,
Philipps-Universit\"at Marburg, D-35032 Marburg, Germany}

%\affiliation{Institute of Semiconductor Physics, 630090 Novosibirsk, Russia}
%\affiliation{Novosibirsk State University, 630090 Novosibirsk, Russia}

%\author {F.~Jansson}
%\affiliation{Department of Physics and Material Sciences Center,
%Philipps-University, D-35032 Marburg, Germany}

\author{K.~Meerholz}
\affiliation{Department f\"ur Chemie, Universit\"at zu K\"oln,
  Luxemburger Stra\ss e 116,
  50939 K\"{o}ln, Germany}

%\author{A.~V.~Dvurechenskii}
%\affiliation{Institute of Semiconductor Physics, 630090 Novosibirsk, Russia}
%\affiliation{Novosibirsk State University, 630090 Novosibirsk, Russia}

% \author{C.~Weisbuch}
% \affiliation{Laboratoire de Physique de la Mati\`ere Condens\'ee,
%   Ecole polytechnique,
%   CNRS,  Universit\'e Paris Saclay, 91128 Palaiseau Cedex, France}
% \affiliation{3Materials Department, University of California, Santa Barbara,
%   California 93106,  USA}
%
% \author{M.~Filoche}
% \affiliation{Laboratoire de Physique de la Mati\`ere Condens\'ee,
%   Ecole polytechnique,
%    CNRS,  Universit\'e Paris Saclay, 91128 Palaiseau Cedex, France}

\author {S.~D.~Baranovskii}
\affiliation{Department of Physics and Material Sciences Center,
  Philipps-Universit\"at Marburg, D-35032 Marburg, Germany}
\affiliation{Department f\"ur Chemie, Universit\"at zu K\"oln,
  Luxemburger Stra\ss e 116,
  50939 K\"{o}ln, Germany}

\date{\today}

%\date{Version as of November 30, 2022}

\begin{abstract}
The current burst in research activities on disordered semiconductors calls for
the development of appropriate theoretical tools that reveal the
features of electron states in random potentials while avoiding the time-consuming
numerical solution of the Schr\"odinger equation.  Among various approaches
suggested so far, the low-pass filter approach of Halperin and Lax (HL)
and the so-called localization landscape technique (LLT)
have received most recognition in the community.
We prove that the HL approach becomes equivalent to the LLT for the specific
case of a Lorentzian filter
when applied to the Schr\"odinger equation with a constant mass.
Advantageously, the low-pass filter approach allows further optimization
beyond the Lorentzian shape. We propose the global HL filter as optimal filter
with only a single length scale, namely, the size of the localized wave packets.
As an application, we design an optimized potential landscape
for a (semi-)classical calculation of the number of strongly localized states
that faithfully reproduce the exact solution
for a random white-noise potential in one dimension.
\end{abstract}

\maketitle   % please do not remove

\section{Introduction}
\label{introduction}

Disordered semiconductors are in the focus of intensive experimental and
theoretical research because these materials are successfully applied
in various devices, and promise many fruitful
applications in the future~\cite{Baranovskii2006Book}.
The term ‘disordered materials’
usually describes amorphous organic and inorganic
solids without perfect crystalline structure.
However, multi-component crystalline semiconductors %with perfect structure
with random atomic occupancies
also belong to the broad class of disordered materials.

Alloying semiconductors is often used to tune material properties, such as
band gaps and effective masses of charge carriers,
depending on the device application.
The price for this tunability is the extra disorder caused
by spatial fluctuations
in the distributions of the material components.
These compositional fluctuations cause spatial fluctuations of the band gap
which in turn create a random potential ${\cal V}(\vec{r}\,)$
acting on electrons and holes~\cite{Masenda2021,WeisbuchNakamura2021}.
The random potential leads to the spatial localization of charge carriers
in electronic states at low energies.

It is desirable to develop theoretical techniques for an appropriate treatment
of such localized states. A complete solution of the Schr\"odinger equation
to obtain all localized states is too time-consuming.
Therefore, theoretical tools to
obtain the essential features of localized states without solving
the Schr\"odinger equation are of extreme value.
One of such tools is the recently introduced `localization landscape theory'
(LLT)~\cite{Arnold2016,LL1_2017,LL2_2017,LL3_2017}.

The LLT can be applied to any wave operator, quantum or classical. In the case
of the Schr\"odinger equation, the localization-landscape potential can be seen
as a regularization of the original potential, and provides an efficient access to the
features of the low-energy
localized states. The random potential ${\cal V }(\vec{r}\,)$
is converted into the `effective potential'
$W(\vec{r}\,) \equiv u^{-1}(\vec{r}\,)$,
where $u(\vec{r}\,)$ is the solution of the LLT equation
\begin{equation}
-\frac{\hbar^2}{2m} \Delta u(\vec{r}\,) + {\cal V}(\vec{r}\,) u(\vec{r}\,)
=1 \; ,
\label{eq:U}
\end{equation}
where $m$ is the carrier mass, $\hbar$ is the reduced Planck constant,
and $\Delta$ is the Laplace operator.
The LLT potential $W(\vec{r}\,)$ is then used to describe, e.g.,
the spatial dependence
of the band edge $E_c(\vec{r}\,)$~\cite{Arnold2016,LL1_2017,LL2_2017,LL3_2017}.

The LLT is widely considered as one of the efficient theoretical approaches
to calculate the local density of states in disordered systems, for instance,
in GaN-based light emitting diodes (LEDs)~\cite{Montoya2021,Tibaldi2021}.
The LLT has been used to simulate the carrier effective
potential fluctuations induced by alloy disorder in InGaN/GaN core-shell
microrods~\cite{Liu2019} and for direct modeling of carrier
transport in multi-quantum-well (QW) LEDs~\cite{Chen2018APL}
and in a type-II superlattice InAs/InAsSb photoconductor
system~\cite{Tsai2020}. The LLT is considered to be a promising approach
to simulate random alloy effects in III-nitride LEDs~\cite{Vito2020},
to connect atomistic
and continuum-based models of III-nitride QWs~\cite{Chaudhuri2021},
and to simulate carrier escape in InGaN/GaN
multiple QW photodetectors~\cite{Chow2020}. Furthermore, the LLT has been applied
to compute the eigenstate localization length at very low energies in
two-dimensional disorder potentials~\cite{Shamailov2021}.
Occasionally, the LLT is considered capable
to reveal errors in the finite-element method software in applications to
alloys \cite{Zhan2021}. It is also used for a three-dimensional modeling of
minority-carrier lateral diffusion length in (In,Ga)N and (Al,Ga)N QWs
and to analysis of light-emission polarization ratio in deep-ultraviolet LEDs by considering
random alloy fluctuations~\cite{Shen2021,Shen2022}.
Moreover, the LLT frequently serves
as a decisive ingredient in quantum-corrected drift-diffusion
models~\cite{Bertazzi2020,Donovan2021,Donovan2022}.

Several attempts to extend and improve
the LLT approach have been made. Chaudhuri
et al.~\cite{Chaudhuri2020} suggested to search for the effective
potential $W(\vec{r}\,)$ replacing Eq.~(\ref{eq:U}), $\widehat{H}u=1$,
by the equation $\widehat{H}^2u=1$. This improves the
convergence of the calculated energies and the robustness of the method against
the chosen integration region for $u$ to obtain the corresponding energies.
Balasubramanian et al.~\cite{Balasubramanian2020} extended the LLT, developed
initially for the single-particle Schr\"odinger operator, to a wide class
of interacting many-body Hamiltonians.
Steinerberger and collaborators~\cite{Steinerberger2017,Steinerberger2021,Lu2022}
suggested to replace the LLT effective potential by
a convolution of the initial random potential with
a kernel function arising from a random Brownian motion.
Altmann et al.~\cite{Altmann2019,Altmann2020} and Jia and
coworkers~\cite{Jia2022} addressed the relation between the LLT and Anderson
localization problem. Harrell and Maltsev adapted the LLT approach for the
eigenfunctions of quantum graphs~\cite{Harrell2020}. Chenn et al.~\cite{Chenn2022}
focused on studying the features of the localized wave functions in the framework of the
LLT. Grubi\v{s}i\'{c} et al.~\cite{Grubisic2021} used the deep neural network model
to study eigenmode localization in the framework of the LLT.
Lemut et al.~\cite{Lemut2020} adapted the LLT to construct the landscape functions
for Dirac fermions. Herviou and Bardarson~\cite{Herviou2020} generalized
the LLT approach to eigenstates at arbitrary energies.

In view of the variety of applications and extensions,
the foundations of the LLT and its accuracy deserve more consideration.
For instance, Comtet and Texier have shown that LLT does not reproduce the energy
spectrum for some simple exactly solvable models,
namely for the Pieces model and for supersymmetric
quantum mechanics~\cite{Comtet2020}.
The group of Witzigmann argued that the LLT is not practical because it is either
not deterministic or requires expensive experimental data. They found that when the
LLT is restricted to a numerically expensive three-dimensional simulation, it does
not perform well in optimization or calibration~\cite{Romer2021}.
In their work, the effect of
compound disorder was investigated with a statistical model. This modelling was
embedded into a multi-scale multi-population carrier-transport simulator to study
its influence on the electronic features~\cite{Romer2021}.
Therefore, the question arises how to justify, and systematically improve, the LLT
for solid-state applications.
In the present work, we elucidate the meaning of the LLT
as a Lorentzian low-pass filter to the random potential, and propose
and justify the Halperin-Lax filter as a generic improvement.

Already in the 1960s, Halperin and Lax~\cite{Halperin1966}
suggested to apply a filter to smooth the random potential.
They  used a variational approach to describe
the shape of the low-energy tail in the density of states
in a random potential.
The absolute square of the wave function serves as their filter function,
whereby the width of the wave function depends on the energy
of the localized state~\cite{Halperin1966}.
Later, Baranovskii and Efros~\cite{Baranovskii1978} addressed the same problem
by a slightly different variational technique and
confirmed the result of Halperin
and Lax for the density of states in a random potential.

However, neither of the two groups considered the individual features
of localized states but restricted themselves to the description
of the tails of the
density of states.
In the present paper, we further develop the idea of
Halperin and Lax~\cite{Halperin1966} to show that
a low-pass filter applied to a random potential
permits to reveal the energy together with the spatial positions
of low-energy localized states in disordered media.

Our paper is organized as follows.
In Sec.~\ref{sec:localization-problem} we briefly
summarize and formalize the problem
of localized states in disordered systems and recall the main ingredients
of the localization landscape
theory~\cite{Arnold2016,LL1_2017,LL2_2017,LL3_2017}.
In Sect.~\ref{sec:LPFapproach} we outline the low-pass filter approach and
show that the localization landscape theory, applied to the Schr\"odinger equation,
is equivalent to
a low-pass filter of Lorentzian form.
In Sec.~\ref{sec:LPFvsvariational}
we show that low-pass filters belongs to the class of variational approaches
to the low-energy localized states, and discuss various local filters
(Gauss, Gauss-Lorentz, Halperin-Lax) that permit to find energy and
position of individual localized states from effective potentials
without solving the Schr\"odinger equation.
Finally, we extend our analysis to define
the Halperin-Lax low-pass filter globally. Comparing with the results
of numerical studies of the localization problem in one dimension, we show
that the global Halperin-Lax filter is superior to the global
Lorentzian low-pass filter used thus far within the framework of the LLT.
Moreover, we construct a landscape potential that can be used for
the semi-classical approximation of the density of states in the band tails.
Concluding remarks are gathered in Sec.~\ref{sec:conclusions}.
In appendix~\ref{app:NofE}, we concisely re-derive the exact integrated particle density
for the case of a random white-noise potential in one dimension.
In appendix~\ref{app:B}, we identify the generic energy and length
scales of the localization problem.

The focus of the LLT and, concomitantly, the focus of the current paper is set on the
features of the temperature-independent effective potential $W(\vec{r}\,)$.
In the LLT,  $W(\vec{r}\,)$ determines the space- and
temperature-dependent electron distribution $n(\vec{r},T)$ that is used
to derive the opto-electronic properties of disordered semiconductors~\cite{LL3_2017}.
However, the effective potential that describes a correct distribution $n(\vec{r},T)$
should also depend on temperature.
In our subsequent paper~\cite{Nenashev_Paper_2}
we present two powerful techniques to access $n(\vec{r},T)$ without solving the
Schr\"odinger equation. First, we derive the density for non-degenerate electrons
by applying the Hamiltonian recursively to random wave functions. Second, we obtain
a temperature-dependent effective potential $W(\vec{r}\,,T)$ from the application of a
universal linear filter to the random potential acting on the charge carriers in disordered
media. Thereby, the full quantum-mechanical problem is reduced to the quasi-classical
description of $n(\vec{r},T)$ in a temperature-dependent effective potential.
Both approaches prove superior to the LLT when we compare our approximate results
for $n(\vec{r},T)$ and for the carrier mobility at elevated temperatures with those from
the exact solution of the Schr\"odinger equation~\cite{Nenashev_Paper_2}.

\section{Localized states in disordered systems}
\label{sec:localization-problem}

We start with a short description of the physical setting
and define a Hamiltonian to model localized states in disordered systems.
Next, we summarize the essence of localization landscape theory.

\subsection{Model}
\label{subsec:Model}

First, we motivate the Hamiltonian used for the description of
quantum mechanical states deeply localized in the band tails.
Second, we define some physical quantities such as the single-particle
density of states
that can be derived from the solution of the Schr\"odinger equation.

\subsubsection{Hamiltonian}
\label{subsubsec:Hamiltonian}

In ideal crystals, there are no impurities and vacancies in the lattice.
As a consequence, Bloch's theorem applies so that
the perfectly periodic lattice potential leads to bands in reciprocal space
with single-particle states that extend over
the whole crystal~\cite{Ashcroft1976}.
In the vicinity of the lower band edge, the dispersion is parabolic so that
we can describe the band states using the kinetic energy
operator in the form
\begin{equation}
  T^{\rm op}(\vec{r}\,)
  = -\frac{\hbar^2}{2m} \Delta = -\frac{\hbar^2}{2m} \sum_{i=1}^d
  \frac{\partial^2 }{\partial r_i^2} \; ,
\end{equation}
where $m$ is the effective mass determined by the curvature of the band
at the $\Gamma$-point; for simplicity, we assume that the band minimum
lies at the $\Gamma$-point in reciprocal space in dimension~$d$.

When a single impurity is added to such a system, an anti-/bound state
forms above/below the upper/lower band edge, provided
the impurity potential is strong enough.
In the presence of a finite density of impurities and other
lattice imperfections,
the band edges smear out and the single-particle states in the band tails are
spatially localized.
When semiconductors are optically excited, the density of charge carriers
is small
and they readily relax into localized states. The trap states and their
properties thus dominate the (hopping) transport
in these materials~\cite{Baranovskii2006Book}.

In alloy semiconductors, the site occupancies randomly vary so that
the charge carriers experience potential fluctuations on an atomic length scale,
$\ell_{\rm pf}\sim 1\, \hbox{\AA}$.
The Hamiltonian to describe charge carriers in the band tails
is thus defined in position space by
\begin{equation}
  {\cal H}_{\rm R}(\vec{r}\,)
  = T^{\rm op}(\vec{r}\,) + {\cal V}_{\rm R}(\vec{r}\,) \; ,
  \label{eq:defHgeneral}
\end{equation}
where ${\cal V}_{\rm R}(\vec{r})$ describes a specific
realization of the fluctuating potential.
To calculate measurable quantities, an average over many realizations~${\rm R}$
must be carried out.

Following Halperin and Lax~\cite{Halperin1966}, we consider a random
potential which arises in the limit of a finite density
of $\delta$-function
scatterers. This is permitted because the strength~${\cal V}$ of
the scattering potential is small compared to the band width
which leads to a typical size of the localized wave packet
$\ell_{\rm wp}\sim 10\, \hbox{\AA}$
that is large compared to the potential fluctuation scale,
$\ell_{\rm wp}\gg \ell_{\rm pf}$.
For an estimate of $\ell_{\rm wp}$ we introduce
\begin{equation}
{\cal T}=   \frac{\hbar^2}{2m\ell_{\rm wp}^2}
  \label{eq:estimatecalT-or-ellwp}
\end{equation}
as the typical average kinetic energy of a localized state.
The estimate~(\ref{eq:estimatecalT-or-ellwp}) implies that,
while the band width is of the order of several electron volts,
the typical average kinetic energy ${\cal T}$ of a localized state
is of the order of several ten milli-electron volts. Algebraic expressions for $\ell_{\rm wp}$ and ${\cal T}$ through the model parameters are given in eq.~(\ref{eq:app-1d-values}).

To be definite, we assume that the potential is
characterized by Gaussian statistics (`white noise'), i.e.,
the potential obeys $ \langle {\cal V}_{\rm R}(\vec{r}\,) \rangle_{\rm R} =0$
with the auto-correlation function~\cite{Halperin1966}
\begin{equation}
 \langle {\cal V}_{\rm R}(\vec{r}\,) {\cal V}_{\rm R}(\vec{r}\,{}^{\prime})
 \rangle_{\rm R} =
 \frac{1}{12}
      {\cal V}^2  (\ell_{\rm pf})^d \delta\left(\vec{r}-\vec{r}\,{}^{\prime}\right)
      \equiv S \delta\left(\vec{r}-\vec{r}\,{}^{\prime}\right)\; ,
 \label{eq:white_noise}
\end{equation}
where $\langle \ldots \rangle_{\rm R}$ indicates
the average over many realizations~${\rm R}$
of the random potential and $S$ is the strength of the interaction.

Only the energetically lowest-lying states are relevant for charge transport
at low carrier concentrations and low temperatures. Therefore,
a third length scale
becomes important, $\ell_{\rm hop}$, for the typically hopping distance
a particle travels
between two localized states. States that are spatially close but
disparate in energy cannot
be accessed due to the lack of activation energy. Therefore, a hopping event
only takes place between localized sites that are close enough in space
and in energy.
Depending on temperature, $\ell_{\rm hop}\sim 100\, \hbox{\AA}$, or larger.
In this work, we shall not discuss transport, and it suffices to state that
localized states that are close in energy are typically separated
in distance by $\ell_{\rm hop}\gg \ell_{\rm wp}$. The precise definition of $\ell_{\rm hop}$ can be found in the monograph by Shklovskii and Efros \cite{Shklovskii1984}.

\subsubsection{Numerical simulations}

Since ${\cal T}$ is the relevant energy scale for the localized states and
$\ell_{\rm wp}$ is the corresponding length scale, we consider the
dimensionless Hamiltonian in $d$~dimensions
\begin{equation}
 \bar{\cal  H}_{\rm R}(\vec{x}\,)= -\sum_{i= 1}^d\frac{\partial^2}{\partial x_i^2}
 + \bar{{\cal V}}_{\rm R}(\vec{x}\,)
 \label{eq:defHreduzed}
\end{equation}
for our numerical simulations,
where we measure lengths in units of $\ell_{\rm wp}$,
$\vec{r}= \ell_{\rm wp}\vec{x}$, and energies
in units of~${\cal T}$, i.e., $E=E/{\cal T}$,
$\bar{{\cal V}}_{\rm R}(\vec{x}\,)=
{\cal V}_{\rm R}(\ell_{\rm wp}\vec{x}\,)/{\cal T}$,
and $\bar{\cal V}= {\cal V}/{\cal T}$.

The calculations can be simplified further when we re-discretize
the Hamiltonian. In this way, we do not have to solve a differential equation
but we are left with the diagonalization of a real symmetric matrix.
Thus, we address the corresponding lattice problem on a $d$-dimensional cube of
edge length~$L$,
\begin{eqnarray}
  \bar{\cal  H}_{\rm R}^{\rm latt}
  &=& -t\sum_{||\vecl-\vecm||= 1}
\left(  |\vecl\rangle \langle \vecm |
+ |\vecm\rangle \langle \vecl |\right) \nonumber \\
&&
  + \sum_{\vecl} \left(2dt+ \bar{\cal V}_{\vecl,{\rm R}}\right)
  |\vecl\rangle \langle \vecl | \; ,
  \label{eq:latticeH3d}
\end{eqnarray}
where $|\vecl\rangle$ denotes a state with a particle on lattice site
$\vecl= (l_1,l_2,\ldots,l_d)$,
and $||\vecl-\vecm||= 1$ implies that the sites $\vecl, \vecm$
are nearest neighbors on a simple cubic lattice of neighboring distance~$a$
in $d$ dimensions.
There are $L_{\rm s}\gg 1$
lattice sites in each direction, i.e.,
$l_j= 1,2,\ldots,L_{\rm s}$, and $L= L_{\rm s}a$.

In one dimension, $\bar{\cal H}_{\rm R}^{\rm latt}$ describes an
$L_{\rm s}\times L_{\rm s}$ matrix with
entries $2t+ \bar{\cal V}_{l,{\rm R}}$ on the diagonal and
$(-t)$ on the secondary diagonal and at its corners.
Matrices with size $L_{\rm s}= {\cal O}(10^4)$
are readily fully diagonalized on present-day computers.

The bare dispersion of the lattice model in $d$~dimensions is
\begin{equation}
\epsilon(\veck)= 2t\sum_{j= 1}^d\left(1-\cos(k_ja)\right)
\end{equation}
where $\veck=(k_1,k_2,\ldots,k_d) =  2\pi/(L_{\rm s}a) (m_1,m_2,\ldots,m_d)$
with integer $-L_{\rm s}/2 \leq m_j< L_{\rm s}/2$ for periodic boundary conditions.
For $\veck \to \veczero$, we have $\epsilon(\veck)\approx ta^2 \veck^2$ which
is the Fourier transform of the Laplace operator in
eq.~(\ref{eq:defHreduzed}) if we identify
\begin{equation}
  t a^2 \equiv 1 \; .
  \label{eq:fixtasfunctionofa}
\end{equation}
The bandwidth must be large compared to the unit energy set by ${\cal T}$
so that we demand $a \ll 1$ for the discretization step to ensure $t \gg 1$.
The microscopic scale in units
of $\ell_{\rm wp}$ is given by $L/L_{\rm s}$
so that we typically choose $a= \ell_{\rm pf}/\ell_{\rm wp}\approx 1/10$.
In this way, the discretization re-introduces the
length-scale for potential fluctuations $\ell_{\rm pf}$
as the microscopic length scale of the localization problem.

The values for the random potential $\bar{\cal V}_{\vecl, {\rm R}}/\bar{\cal V}_L$
are chosen with equal probability from the interval
$I= [-1/2,1/2]$ (box distribution). Then, averaging over
many realizations~${\rm R}$ gives
\begin{eqnarray}
  \langle \bar{\cal V}_{\vecl,{\rm R}}\rangle_{\rm R} &=& \bar{\cal V}_L
  \int_{-1/2}^{1/2} {\rm d} p \, p = 0 \; , \nonumber \\
  \langle \bar{\cal V}_{\vecl,{\rm R}}\bar{\cal V}_{\vecm,{\rm R}}
  \rangle_{\rm R} &=& \delta_{\vecl,\vecm} \bar{\cal V}_L^2
  \int_{-1/2}^{1/2} {\rm d} p \, p^2 = \frac{\bar{\cal V}_L^2}{12}\delta_{\vecl,\vecm}
  \label{eq:autocorrelationlattice}
  \end{eqnarray}
because $\bar{\cal V}_{\vecl,{\rm R}}/\bar{\cal V}_L$
has the value~$p$ with equal probability
$w(p)= 1$ in the interval~$I$.

Eq.~(\ref{eq:autocorrelationlattice}) is the lattice version of
eq.~(\ref{eq:white_noise}).
Correspondingly, the dimensionless strength~$S_L$ of the potential is given by
\begin{equation}
S_L= \frac{\bar{\cal V}_L^2 a^d}{12}
\end{equation}
or
\begin{equation}
  \bar{\cal V}_L= \sqrt{\frac{12 S_L}{a^d}}
  \label{eq:calVandS}
\end{equation}
for the prefactor $\bar{\cal V}_L$ in the potential as a function of the
strength~$S_L$ and of the discretization~$a$.
The spectrum at low energies scales with the impurity
interaction strength~$S_L$.

In our numerical simulations in one dimension,
we choose $a= 0.1$, i.e., the potential fluctuates on the scale
$\ell_{\rm pf}= a \ell_{\rm wp}= \ell_{\rm wp}/10$.
The half-width at half maximum of the wave packets in position space
is of the order $\ell_{\rm wp}$ when we set $S_L\equiv 1$ so that the
typical energies of the bound states are of order ${\cal T}\equiv 1$.
On average, the energy and the width
of the wave packets are proportional
to $S_L^{\nu}$ with $\nu_e^{d=1}=2/3$ for the energy and
$\nu_{\rm w}^{d=1}=-1/3$ for the width, see appendix~\ref{app:B}.

\subsubsection{Physical quantities}
\label{subsubsec:observables}

The solutions of the Schr\"odinger equation
\begin{equation}
  {\cal H}_{\rm R}(\vec{r}\,) \psi(\vec{r}\,) = E \psi(\vec{r}\,)
  \label{eq:Seq3d}
\end{equation}
with the Hamiltonian
from eq.~(\ref{eq:defHgeneral}) provide the spectrum $\{ E_{l,{\rm R}}\}$
and the corresponding eigenfunctions $\{ \psi_{l,{\rm R}}(\vec{r}\,)\}$
from which several
physical quantities of interest can be deduced.

The single-particle density of states is defined by
\begin{equation}
  \rho(E) =\frac{1}{L} \langle \sum_l \delta(E-E_{l,{\rm R}})\rangle_{\rm R} \; .
  \label{eq:dosdef}
\end{equation}
It permits to derive the (average) Fermi level $E_{\rm F}$
for given particle density~$n$,
\begin{equation}
n= \frac{N}{L}= \int_{-\infty}^{E_{\rm F}} {\rm d} E \rho(E) \; .
\end{equation}
The local particle density at temperature~$T$
for a given realization of the impurity potential
is given by ($k_{\rm B}\equiv 1$)
\begin{equation}
  n_{\rm R}(\vec{r},\mu_{\rm R},T) = \sum_l e^{-(E_{l,{\rm R}}-\mu_{\rm R})/T}
  |\psi_{l,{\rm R}}(\vec{r}\,)|^2
\end{equation}
with the normalization
\begin{equation}
\int {\rm d}^3r\, n_{\rm R}(\vec{r},\mu_{\rm R},T) =1
\end{equation}
that fixes the chemical potential $\mu_{\rm R}\equiv \mu_{\rm R}(T)$.

The two-point correlation function remains non-trivial after averaging,
\begin{eqnarray}
  C(\vec{r}-\vec{r}\,{}^{\prime},T)
  &\equiv &
  C(\vec{r},\vec{r}\,{}^{\prime},T)
   \nonumber \\
  &=&
   \langle \sum_{l_1,l_2}
   e^{(E_{{l_1,{\rm R}}}-\mu_{\rm R})/T}
  |\psi_{l_1,{\rm R}}(\vec{r}\,)|^2 \nonumber \\
  && \hphantom{\langle \sum_{l_1,l_2}}
   e^{(E_{{l_2,{\rm R}}}-\mu_{\rm R})/T}
  |\psi_{l_2,{\rm R}}(\vec{r}\,{}^{\prime})|^2
  \rangle_{\rm R} \nonumber \\
  &&  - n(\vec{r},T)n(\vec{r}\,{}^{\prime},T) \; , \nonumber \\
  n(\vec{r},T) \equiv n(T) &= &
  \langle n_{\rm R}(\vec{r},\mu_{\rm R},T)\rangle_{\rm R}
  \; .
\end{eqnarray}
The two-point correlation function permits to identify the typical
distance between occupied sites at low temperature, i.e., it
can be used to determine~$l_{\rm hop}$.

Another quantity of interest is the spatially resolved density-of-states
for a given realization of the impurity potential
\begin{equation}
  \rho_{\rm R}(E,\vec{r}\,) = \sum_l |\psi_{l,{\rm R}}(\vec{r}\,)|^2
  \delta(E_{l,{\rm R}}-E)
\end{equation}
because it contains the information
where states are localized in position space
that have the energy~$E$.

\subsection{Localization landscape theory}
\label{subsec:LLTsketch}

The calculation of the density of states and other physical
quantities requires the full spectrum for many realizations.
Complete numerical calculations can be carried out
for fairly large one-dimensional systems where
also exact solutions for some quantities are available for comparison.
Apparently, this is not
feasible for large systems in three dimensions, in particular when
these microscopic calculations are part of larger
program packages that aim to describe transport on mesoscopic
length scales self-consistently.
Therefore, approximate theories must be developed.

One successful approach is the localization
landscape theory (LLT)~\cite{Arnold2016,LL1_2017,LL2_2017,LL3_2017}.
It starts from the idea that the deeply localized states can be described
by some approximate
potential $W_{\rm R }(\vec{r}\,)$
that is much smoother than the white-noise potential
and describes the landscape that the localized states actually encounter in
transport. Moreover, the smooth localization
landscape $W_{\rm R}(\vec{r}\,)$
permits to calculate the density of states and the local density of states
from semi-classical expressions~\cite{Arnold2016,LL1_2017,LL2_2017,LL3_2017}.

In LLT, the landscape potential $W_{\rm R}(\vec{r}\,)=1/u_{\rm R}(\vec{r}\,)$
is determined from the solution of a Poisson-type equation,
\begin{equation}
{\cal H}_{\rm R} u_{\rm R}(\vec{r}\,) = 1
\end{equation}
that has to be solved only once for a given realization
of the disorder potential, see eq.~(\ref{eq:U}).
Thus, averaging over many realizations does not pose an insurmountable problem,
and the solution of the microscopic problem can be incorporated into
mesoscopic transport equations. It turns out that it is favorable to apply
a Gaussian smoothing to the white-noise potential, i.e., the LLT is not
directly applied to ${\cal V}_{\rm R}(\vec{r}\,)$ but to
\begin{eqnarray}
  {\cal V}_{\rm R, av}(\vec{r}\,)&=& \frac{1}{D}
  \int_{-\infty}^{\infty} {\rm d}^3 r'
   \exp\left(-\frac{(\vec{r}-\vec{r}\,{}^{\prime})^2}{2\sigma^2}\right)
   {\cal V}_{\rm R}(\vec{r}\,{}^{\prime})
   \nonumber \\
   D&=&
   \int_{-\infty}^{\infty} {\rm d}^3 r'
   \exp\left(-\frac{(\vec{r}-\vec{r}\,{}^{\prime})^2}{2\sigma^2}\right)
  \;,
\label{eq:random_potential_correlated}
\end{eqnarray}
where $\sigma$ is the spatial scale of the Gaussian averaging.
With these modifications and extensions,
the LLT was successfully applied to, e.g., mesoscopic
carrier transport and recombination in light emitting diodes~\cite{LL3_2017}.
Therefore, it is worthwhile to elaborate on the foundations of LLT
in more detail.

\section{Low-pass filter approach}
\label{sec:LPFapproach}

In this section we show that the LLT, when applied to the Schr\"odinger equation
with a constant mass, is a special version of a more general
low-pass filter approach to localized states at low energies.

\subsection{Derivation}
\label{subsec:derivation}

For simplicity and illustration, we study the one-dimensional Schr\"o\-din\-ger equation.
The resulting structure of the energetically lowest-lying states
motivates the introduction
of low-pass filters to the localization problem.

\subsubsection{Motivation}

\begin{figure}[t]%
  \begin{center}
    \begin{tabular}[t]{@{}l@{}}
  (a) \\
  \includegraphics[width=\linewidth]{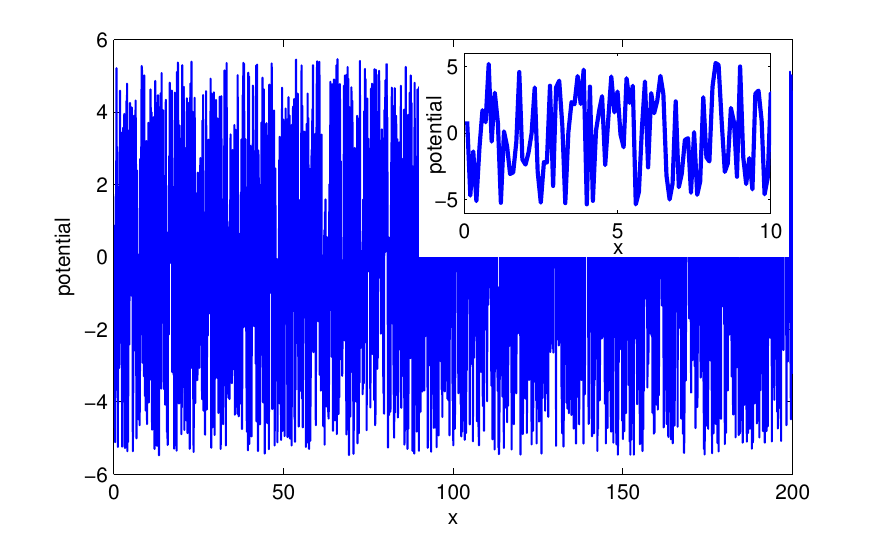}\\
 % \hspace{8cm}\vspace*{6cm}\\
  (b) \\
  \includegraphics[width=\linewidth]{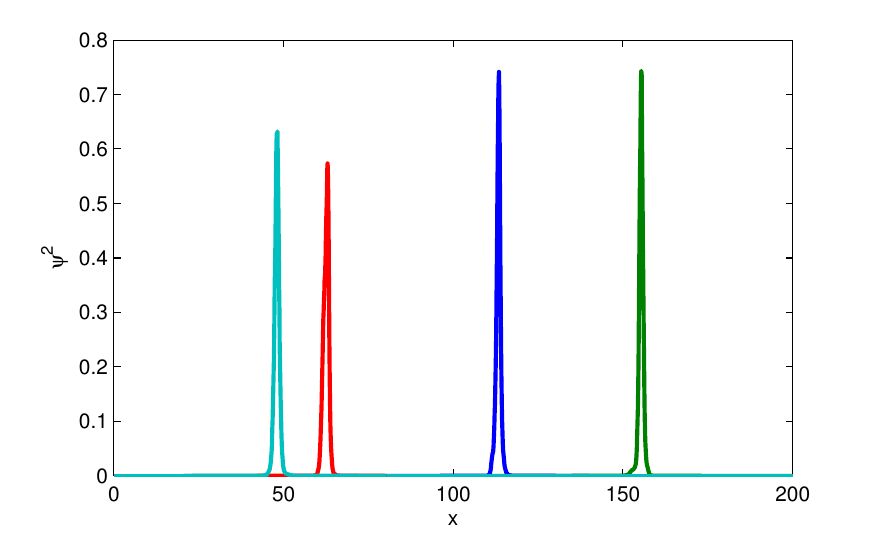}
 % \hspace{8cm}\vspace*{6cm}
    \end{tabular}
    \end{center}
\caption{(a) Realization for the white-noise impurity potential
  on a one-dimensional
  strip with $L_{\rm s}= 2.000$ sites for $L= 200$ ($a= 0.1$).
  (b) Occupation probabilities $|\psi_{l,{\rm R}}(n)|^2$ for
  the four lowest-energy eigenstates ($l= 1,2,3,4$).
  At impurity strength $S_L=1$, the energies, positions, and
  standard deviations of the four states are
  $e=E/{\cal T}=(-1.44,-1.17,-1.10,-0.88)$,
  $\bar{x}=(113,155,63,48)$,
  $\Delta x=\sqrt{\bar{x^2}-\bar{x}^2}
  =(0.69,0.71,0.79,0.76)$.\label{fig:potentialandstates}}
\end{figure}

We address the one-dimensional Schr\"o\-din\-ger equation
\begin{eqnarray}
  \bar{\cal H}_{\rm R}^{\rm 1d} |\psi_{l,{\rm R}}\rangle
  &= & e_{l,{\rm R}} | \psi\rangle \; , \nonumber \\
  |\psi_{l, {\rm R}}\rangle &=& \sum_{n= 1}^{L_{\rm s}} \psi_{l,{\rm R}}(n) |n\rangle
  \label{eq:specific1dHamilt}
\end{eqnarray}
with
\begin{eqnarray}
  \bar{\cal H}_{\rm R}^{\rm 1d} &=& -t \sum_{n= 1}^{L_{\rm s}}
  \left(  |n\rangle \langle n+ 1 |
+ |n+ 1\rangle \langle n |\right) \nonumber \\
&&  + \sum_{n= 1}^{L_{\rm s}} \left(2t+ \bar{\cal V}_{n,{\rm R}}\right)
  |n\rangle \langle n |
  \label{eq:latticeH1d}
\end{eqnarray}
for periodic boundary conditions, $|n\rangle \equiv |n+ L_{\rm s}\rangle$
and $t= 1/a^2$, see eq.~(\ref{eq:fixtasfunctionofa}), for $a= 0.1$.
We specify a realization of the disorder potential~${\rm R}$, i.e.,
we choose $L_{\rm s}$ random numbers from the interval $I= [-1/2,1/2]$
and multiply it with the amplitude $\bar{\cal V}_L= \sqrt{12/a}$ for $S_L=1$ from
eq.~(\ref{eq:calVandS}).
A realization of the disorder potential is shown
in Fig.~\ref{fig:potentialandstates}(a) together
with the occupation probabilities
in position space $|\psi_{l,{\rm R}}(n)|^2$ for the four lowest-lying states
in Fig.~\ref{fig:potentialandstates}(b).
The amplitudes are real for bound states in one dimension.

As seen from the figure, the potential oscillates
from lattice site to lattice site, on the length scale
$\ell_{\rm pf}=a \ell_{\rm wp}= \ell_{\rm wp}/10$.
In contrast, the wave functions for the low-lying localized states are
extended over the length scale $\ell_{\rm wp} \gg \ell_{\rm pf}$.
This behavior is readily understood from quantum mechanics.
Classical particles at low energy would localize
at the minima of the disorder potential.
For quantum mechanical particles this would imply a very high kinetic energy
so that the uncertainty principle leads to a spread-out of the wave function
in position space to reach a low total energy.
The barriers between adjacent sites are high but they are also narrow
so that tunneling between sites permits wave functions that spread over
many neighboring sites.

The spreading of the ground-state wave function in position space
over distances $\ell_{\rm wp}$ implies an equally localized
wave packet in reciprocal space,
\begin{equation}
|\psi_{l,{\rm R}}\rangle = \sum_k \tilde{\psi}_{l,{\rm R}}(k) |k\rangle\; ,
\end{equation}
where
\begin{equation}
  |  k\rangle= \sqrt{\frac{1}{L_{\rm s}}} \sum_{n= 1}^{L_{\rm s}}
  e^{{\rm i} k n a} |n\rangle
\end{equation}
are the eigenstates of the kinetic energy operator
with wave number $k= 2\pi m_k/(L_{\rm s}a)$,
$m_k= -L_{\rm s}/2, \ldots, L_{\rm s}/2-1$,
and energy $\epsilon(k)= 2t(1-\cos(ka))$.

\begin{figure}[b]
  \begin{center}
    \includegraphics[width=\linewidth]{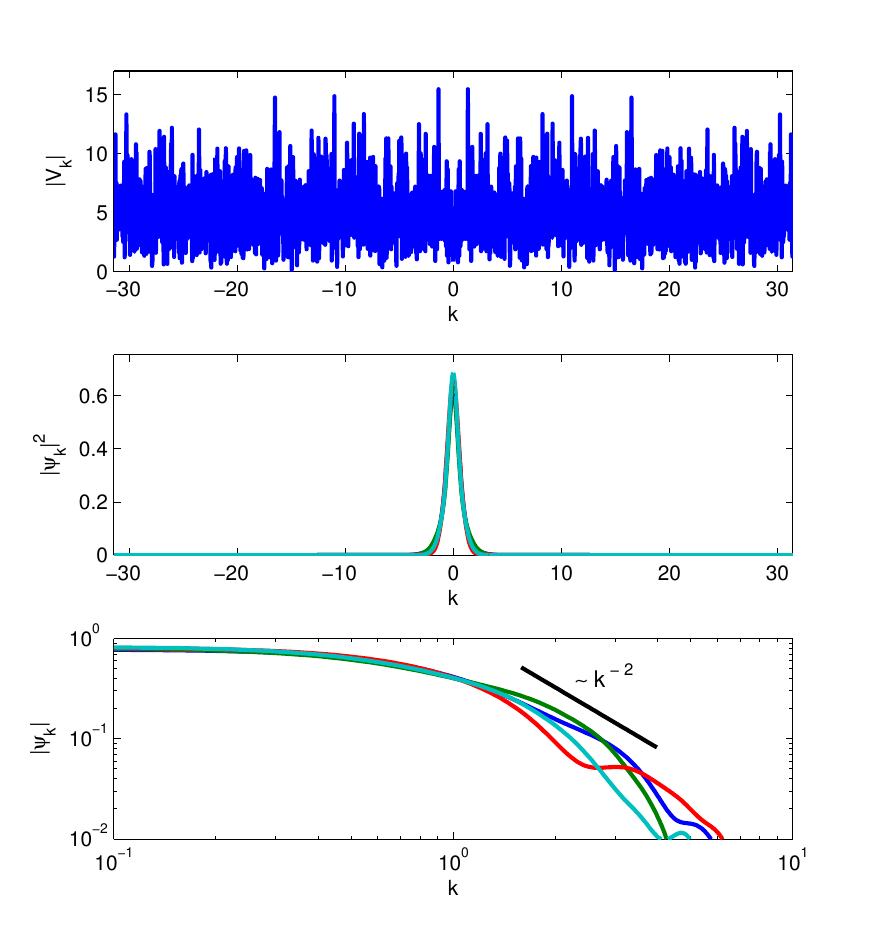}
% \vspace*{2cm}
     \end{center}
  \caption{Wave function intensities in momentum space $|\tilde{\psi}_{l,{\rm R}}(k)|^2$
    for the four states with lowest energy in the potential landscape
    of Fig.~\protect\ref{fig:potentialandstates}, shown in momentum space
    in the top part of the figure.
        Note that the Brillouin zone boundaries are
    at $\pi/a=10\pi$.
    The bottom part of the figure shows $|\tilde{\psi}_{l,{\rm R}}(k)|$ on a
    log-log scale.\label{fig:wavefunction-kspace}}
\end{figure}

The amplitude squares $|\tilde{\psi}_{1,{\rm R}}(k)|^2$ for the ground state
of the Hamiltonian~(\ref{eq:specific1dHamilt})
are shown in Fig.~\ref{fig:wavefunction-kspace}; the specific realization
of the disorder potential is the same as
shown in Fig.~\ref{fig:potentialandstates}(a).
It is seen that a typical low-energy
wave packet spreads over a distance $1/\ell_{\rm wp}$
in reciprocal space, and vanishes algebraically for large~$k$,
\begin{equation}
  \tilde{\psi}(k) \propto \frac{1}{(k\ell_{\rm wp})^2}
  \label{eq:psidecayinFourierspace}
\end{equation}
in one dimension.

This behavior is readily shown to be generic for low-energy states
of a one-dimensional Hamiltonian with site-diagonal disorder.
We transform the Schr\"odinger equation~(\ref{eq:Seq3d})
in one dimension to reciprocal space,
\begin{equation}
  \frac{\hbar^2k^2}{2m} \tilde{\psi}(k)
  + \int \frac{{\rm d} q}{2\pi} \tilde{\cal V}_{\rm R}(k-q)\tilde{\psi}(q)
  = E \tilde{\psi}(k)
  \label{eq:SeqFT}
\end{equation}
with
\begin{equation}
  \tilde{\cal V}_{\rm R}(q) =
  \int_0^L {\rm d } x e^{-{\rm i}q x} {\cal V}_{\rm R}(x)  \; .
  \label{eq:VRFouriertransformed}
\end{equation}
{}From eq.~(\ref{eq:SeqFT}), we see that
\begin{equation}
  \tilde{\psi}(k) = \frac{1}{\hbar^2k^2/(2m)-E+ {\rm i}\eta}
  \int \frac{{\rm d} q}{2\pi}
  \tilde{\cal V}_{\rm R}(k-q)\tilde{\psi}(q) \; ,
  \label{eq:FTusefullargek}
\end{equation}
where $\eta= 0^+ $ to ensure a proper analytic behavior around the pole.
Eq.~(\ref{eq:FTusefullargek}) shows that, for small~$E$ and large~$k$,
we reproduce the empirically observed behavior
in eq.~(\ref{eq:psidecayinFourierspace})
because, for a representative $\tilde{\cal V}_{\rm R}(k)$,
the integral becomes independent of~$k$
when $|k|$ becomes large.

\subsubsection{Definition of the low-pass filter}

Eq.~(\ref{eq:FTusefullargek}) shows that for small~$k$
only small $(k-q)$ values contribute
to the integral because $k$ and $q$ are both small. Therefore,
only small Fourier components of the random potential
in eq.~(\ref{eq:VRFouriertransformed}) are important.

This observation constitutes the basis of the low-pass filter approach.
The effective random potential
\begin{equation}
  \tilde{W}_{\rm R}(k) = \tilde{\Gamma}(k)
  \tilde{\cal V}_{\rm R}(k)
\end{equation}
leads to the same solution of the Schr\"odinger equation~(\ref{eq:SeqFT})
when we apply the low-pass filter $\tilde{\Gamma}(k)$ that fulfills the
conditions
\begin{eqnarray}
  \tilde{\Gamma}(k\ll 1/\ell) &\approx & 1 \; ,\label{eq:normofGamma} \\
  \tilde{\Gamma}(k \gg 1/\ell) &\approx & \frac{1}{(k\ell)^2} \; ,
  \label{eq:conditionsonGammatildek}
\end{eqnarray}
where $\ell$ is a characteristic length scale that remains to be adjusted.
Eq.~(\ref{eq:normofGamma}) implies that a constant shift in $\bar{\cal V}_{\rm R}(x)$
leads to the same constant shift in $\bar{W}_{\rm R}(x)$.

The effective potential in position space is given by
the convolution of the original potential and a weight function,
\begin{equation}
  W_{\rm R} (x) = \int_0^L {\rm d} x' \Gamma(x-x') {\cal V}_{\rm R}(x')
  \label{eq:convolution}
\end{equation}
with
\begin{equation}
\Gamma(x) = \int \frac{{\rm d} q}{2\pi} e^{{\rm i}qx}\tilde{\Gamma}(q) \; .
\end{equation}
This shows that the oscillations of the white-noise potential are
smoothed by the filter function $\Gamma(x)$.

A particularly simple choice for $\tilde{\Gamma}(k)$ that fulfills
eq.~(\ref{eq:conditionsonGammatildek}) is
a Lorentzian low-pass filter,
\begin{equation}
  \tilde{\Gamma}^{\rm L}(k)
= \frac{1}{1+ (k\ell_{\rm L})^2} \; .
\label{eq:Lorentz_low_pass_filter}
\end{equation}
In this way, the effective potential in position space is given by
the convolution of the original potential and an exponential
weight function,
\begin{equation}
  W_{\rm R}^{\rm L}(x)
  =  \int_0^L \frac{{\rm d} x'}{2\ell_{\rm L}}
  e^{-|x-x'|/\ell_{\rm L}} {\cal V}_{\rm R}(x') \; .
  \label{eq:convolutionLorentz}
\end{equation}
Apparently, the original potential is averaged
over the length scale~$\ell_{\rm L}$.

\subsection{Relation to localization landscape theory}
\label{subsec:LPFversusLLT}

Apart from its apparent simplicity,
the Lorentzian low-pass filter offers
yet another advantage:
it permits to make close contact with
the LLT~\cite{Arnold2016,LL1_2017,LL2_2017,LL3_2017}.

\subsubsection{LLT as a low-pass filter}

We note
\begin{eqnarray}
  -\frac{{\rm d}^2}{{\rm d} x^2}
  W_{\rm R}^{\rm L}(x)
  &=&  \int \frac{{\rm d} k}{2\pi}
    \frac{k^2}{1+ (k\ell_{\rm L})^2} e^{{\rm i}k x} \tilde{\cal V}_{\rm R}(k)
 \nonumber   \\
    &=&  \frac{1}{\ell_{\rm L}^2}
    \int \frac{{\rm d} k}{2\pi}
    e^{{\rm i}k x} \tilde{\cal V}_{\rm R}(k) \nonumber \\
&&    - \frac{1}{\ell_{\rm L}^2} \int \frac{{\rm d} k}{2\pi}
    \frac{1}{1+ (k\ell_{\rm L})^2} e^{{\rm i}k x} \tilde{\cal V}_{\rm R}(k)
    \nonumber \\
    \\
    &=&  \frac{1}{\ell_{\rm L}^2}\left( {\cal V}_{\rm R}(x)
    - W_{\rm R}^{\rm L}(x)\right) \; .
\end{eqnarray}
Therefore, we obtain the defining equation for the effective potential
with a Lorentzian low-pass filter in position space
\begin{equation}
  \ell_{\rm L}^2 \frac{{\rm d}^2}{{\rm d} x^2}
  W_{\rm R}^{\rm L}(x) +  {\cal V}_{\rm R}(x)=  W_{\rm R}^{\rm L}(x)
  \; .
  \label{eq:defequationpositionspace}
\end{equation}
In LLT, the defining equation for the effective potential also only involves
the potential itself,
\begin{equation}
  -\frac{\hbar^2}{2m} \frac{{\rm d}^2}{{\rm d} x^2}
  u_{\rm R}(x) + {\cal V}_{\rm R}(x) u_{\rm R}(x) = 1
  \label{eq:LLTequation}
\end{equation}
or, with
\begin{equation}
 W_{\rm R} (x) = \frac{1}{u_{\rm R}(x)} \; ,
\end{equation}
we find the LLT equation
\begin{equation}
  W_{\rm R}(x) \left(-\frac{\hbar^2}{2m} \frac{{\rm d}^2}{{\rm d} x^2}
  \right)\left(\frac{1}{W_{\rm R}(x)}\right)
  + {\cal V}_{\rm R}(x) = W_{\rm R}(x) \; ,
  \label{eq:LLTalmostmyequation}
\end{equation}
which is very similar to the
Lorentzian low-pass filter equation~(\ref{eq:defequationpositionspace}).
We note that, for all practical applications,
\begin{equation}
  u_{\rm R}(x) = c + \delta u_{\rm R}(x)
  \label{eq:introduce-c}
\end{equation}
with
\begin{equation}
|\delta u_{\rm R}(x)| \ll c
\end{equation}
holds. Therefore,
in linear approximation, we have
\begin{equation}
  W_{\rm R}(x)
    = \frac{1}{u_{\rm R}(x)} \approx \frac{1}{c}\left(1-\frac{1}{c}
  \delta u_{\rm R}(x)\right)
\end{equation}
so that
\begin{eqnarray}
  \frac{{\rm d}^2}{{\rm d} x^2}
  W_{\rm R}(x)
  &\approx & -\frac{1}{c^2}\frac{{\rm d}^2}{{\rm d} x^2}
  \delta u_{\rm R}(x) \nonumber \\
  &=& -\frac{1}{c^2} \frac{{\rm d}^2}{{\rm d} x^2} u_{\rm R}(x)\nonumber \\
  &=& -\frac{1}{c^2} \frac{{\rm d}^2}{{\rm d} x^2}
  \left(\frac{1}{W_{\rm R}(x)}\right)
  \; .
\end{eqnarray}
When inserted into eq.~(\ref{eq:LLTalmostmyequation}), we find
to linear order
\begin{equation}
\frac{1}{c} \left(-\frac{\hbar^2}{2m} \frac{{\rm d}^2}{{\rm d} x^2}
  \right)\left(-c^ 2 W_{\rm R}(x)\right)
  + {\cal V}_{\rm R}(x) = W_{\rm R}(x)\; ,
  \label{eq:LLTmyequation}
\end{equation}
which reduces to eq.~(\ref{eq:defequationpositionspace})
when we identify
\begin{eqnarray}
  W_{\rm R}(x)= W_{\rm R}^{\rm L}(x)\; , \label{eq:W_equalW_LLT}\\
  \frac{1}{c}= \frac{\hbar^2}{2m\ell_{\rm L}^2} \; .
  \label{eq:candell}
\end{eqnarray}
Note that, since $\delta u_{\rm R}(x)$ is evenly distributed around zero,
we have
\begin{equation}
  \frac{1}{L} \int_0^L {\rm d} x W_{\rm R}(x) \approx
  \frac{1}{L} \int_0^L {\rm d} x \frac{1}{c}
  \left( 1-\frac{1}{c} \delta u_{\rm R}(x)\right)  = \frac{1}{c} \; .
  \label{eq:getcfromWaverage}
\end{equation}
Therefore, the solution of the LLT equation~(\ref{eq:LLTequation})
gives~$c$ and thus $\ell_{\rm L}$ via eq.~(\ref{eq:candell}).
It means that the Lorentz-filtered potential is equivalent to the LLT
potential when the Lorentz parameter $\ell_{\rm L}$
is determined from eq.~(\ref{eq:candell}),
where $c^{-1}$ is the mean value of $W_{\rm R}(x)$.

\begin{figure}[t]
  \includegraphics[width=\linewidth]{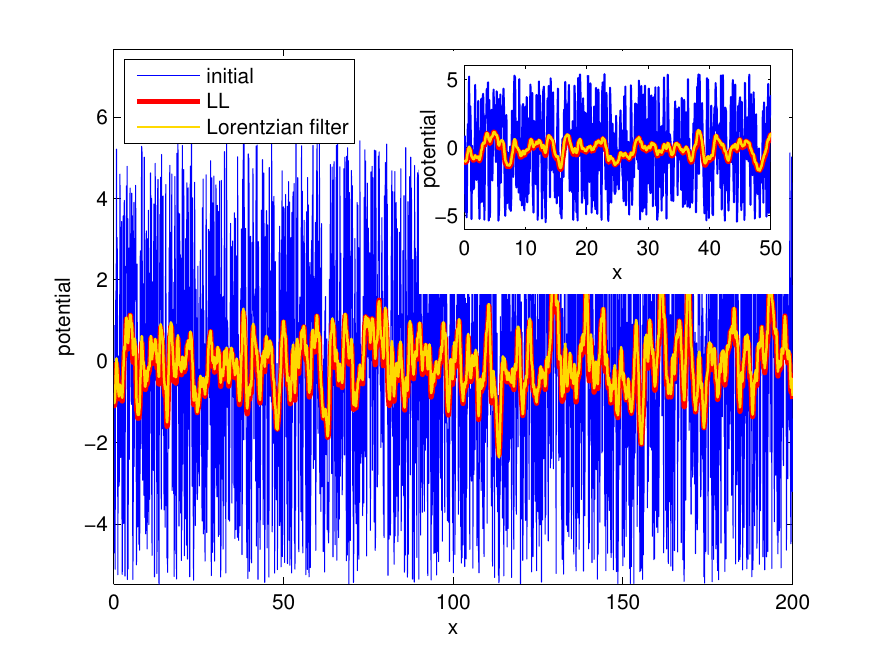}
  \caption{Comparison between LLT and a global Lorentzian filter.
    The random potential ${\cal V}_{\rm R}(x)$
    of Fig.~\protect\ref{fig:potentialandstates}(a) is shown as a blue line.
    The red line is the result of LLT, $W_{\rm R}(x)$,
    from the solution of Eq.~(\ref{eq:LLTequation}).
    The yellow line shows the potential $W_{\rm R}^{\rm L}(x)$
    using a Lorentzian filter, see eq.~(\ref{eq:convolutionLorentz}),
    where $\ell_{\rm L}=0.436\, \ell_{\rm wp}$
    is calculated from eqs.~(\protect\ref{eq:candell})
    and~(\protect\ref{eq:getcfromWaverage}).\label{fig:Lorentzian}}
\end{figure}

\subsubsection{Visualization}

In order to visualize the relation between LLT and the Lorentzian
low-pass filter, we re-consider the one-dimensional random potential shown in
Fig.~\ref{fig:potentialandstates}(a).
The LLT requires the solution of the Poisson-type
equation~(\ref{eq:LLTequation}). The spatial average of the solution
$u_{\rm R}(x)$ defines~$c$, see eq.~(\ref{eq:introduce-c}), which
in turn gives the parameter~$\ell_{\rm L}$ in the Lorentzian low-pass filter
using eq.~(\ref{eq:candell}).

In Fig.~\ref{fig:Lorentzian} we compare the effective potentials from
LLT and from the Lorentzian low-pass filter.
Apparently, the agreement is very good. This one-dimensional
example illustrates that LLT and a Lorentzian low-pass filter
applied to the white-noise potential lead to equivalent results.

\subsubsection{Advantages of the low-pass filer approach}

The low-pass filter approach
offers several advantages.
(i) There are no conceptual problems whereas
some gradient terms in the Schr\"odinger equations are ignored in the
motivation of the LLT.
(ii) The physical picture of a coarse-grained potential
  becomes more obvious, see eq.~(\ref{eq:convolution}).
  The coarse-graining is permitted because the low-energy wave functions
  themselves decay algebraically as a function of momentum.
(iii) The low-pass filter is not restricted to a Lorentzian shape
but the filter functions can be adapted
for specific applications.

In the next section we address the improvement of the low-pass filter
beyond the Lorentzian shape.

\section{Variational approaches as low-pass filters}
\label{sec:LPFvsvariational}

In the last section, the length-scale parameter~$\ell$ in the Lorentzian
low-pass filter was fixed by comparison with LLT. In general,
however, one would prefer a criterion to optimize not only
that length scale but to find the best filter function according to
some physical criterion. Variational theories are best suited
for this task because they optimize physically motivated
cost functions.

The variational approach to the localization problem is far from new.
Indeed, almost six decades ago,
Halperin and Lax~\cite{Halperin1966} used this approach to determine
variationally the average density of states $\rho_{\rm var}(E)$;
the result was quantitatively improved by including
higher-order corrections~\cite{Halperin1967}.
For a Green function approach to white-noise disorder,
see Zittartz and Langer~\cite{PhysRev.148.741}.

\subsection{Variational theory for localized states}
\label{subsec:variational-theory}

\subsubsection{Ritz variational principle}

The Ritz variational principle states that
any normalized
single-particle wave function $\varphi(\vec{r}\,) $
provides an upper bound to the exact ground-state energy,
\begin{equation}
  E_{\rm R}^{\rm var}(\left\{\varphi\right\})
  =  \int {\rm d}^3 r
  \varphi^*(\vec{r}\,)  {\cal H}_{\rm R}(\vec{r}\,) \varphi(\vec{r}\,) \geq
  E_{{\rm R},0 }\; ,
  \label{eq:RitzVP}
\end{equation}
where $E_{{\rm R},0}$ is the exact ground-state energy
for the realization~${\rm R}$.
For a given variational state
one has to calculate the kinetic energy,
\begin{equation}
t_{\rm var}(\left\{\varphi(\vec{r}\,)\right\}) = \int {\rm d}^3 r
  \varphi^*(\vec{r}\,)  T^{\rm op}(\vec{r}\,) \varphi(\vec{r}\,) \; ,
\label{eq:Tvariational}
\end{equation}
and the potential energy,
\begin{equation}
  v_{\rm var}(\left\{\varphi(\vec{r}\,)\right\})
  = \int {\rm d}^3 r
  |\varphi(\vec{r}\,)|^2
  {\cal V}_{\rm R}(\vec{r}\,) \; .
\label{eq:calVvariational}
\end{equation}
To carry out the Ritz variational optimization, the variational
states must be further specified.

\subsubsection{Variational localized ground state}

For the localization problem,
we assume that low-energy states are localized around some
$\vec{r}\,{}^{\prime}$~\cite{Halperin1966},
\begin{equation}
  \varphi(\vec{r}\,)\equiv
  \varphi_{\vecalpha}(\vec{r}\,{}^{\prime}-\vec{r}\,) \; .
\end{equation}
Note that the variational state depends on $\vec{r}\,{}^{\prime}$ and
further internal parameters $\vecalpha=(\alpha_1,\alpha_2, \ldots,)$
that are to be determined variationally.
It is assumed that the wave function vanishes fast
for $|\vec{r}-\vec{r}\,{}^{\prime}|\gg \ell_{\rm wp}$.

The variational kinetic energy is independent of $\vec{r}\,{}^{\prime}$ because
it only induces a constant shift in position space in the wave function. Therefore, the
minimization of the variational energy
$E_{\rm R}^{\rm var}(\left\{\varphi_{\vecalpha}(\vec{r}\,{}^{\prime})\right\})$
in eq.~(\ref{eq:RitzVP}) with respect to $\vec{r}\,{}^{\prime}$
leads to ($i=1,\ldots,d$)
\begin{equation}
  \frac{\partial}{\partial r_i^{\prime}}
  \int {\rm d}^3r \left|\varphi_{\vecalpha}(\vec{r}\,{}^{\prime}-\vec{r}\,)\right|^2
    {\cal V}_{\rm R}(\vec{r}\,) =0 \; .
\end{equation}
In other word, the wave function is localized around the minimum of the
$\varphi$-potential
\begin{equation}
  W_{\rm R}^{\varphi}(\vec{r},\vecalpha)
  =
  \int {\rm d}^3r' \left|\varphi_{\vecalpha}(\vec{r}-\vec{r}\,{}^{\prime})\right|^2
    {\cal V}_{\rm R}(\vec{r}\,{}^{\prime}) \; .
  \label{eq:convolution3dvariational}
\end{equation}
Note that the variational state and energy remain to be optimized with respect to
the variational parameters $\vecalpha$ in $\varphi$, i.e.,
the total variational energy
\begin{equation}
  E_{\rm R}^{\varphi}(\vec{r},\vecalpha) = t_{\rm R}^{\varphi}(\vecalpha) +
  W_{\rm R}^{\varphi}(\vec{r},\vecalpha)
\end{equation}
remains to be optimized with respect to $\vecalpha$.
In this way, position, shape, and energy of
the ground state of the Hamiltonian ${\cal H}_{\rm R}(\vec{r}\,)$ are
determined variationally.

Apparently, we may {\rm interpret\/}
$W_{\rm R}^{\varphi}(\vec{r},\vecalpha)$ at the optimized
values as the effective potential landscape in the vicinity
of $\vec{r}\,{}^{\prime}$ where the particle is localized in the ground state.

\subsection{Local low-pass filters}
\label{subsec:local-low-pass-filters}

\subsubsection{Concept}

In general, the Ritz variational principle is restricted
to find an upper bound for
the ground-state energy. For the localization problem,
it can be used to detect
many more energetically low-lying states
because these are typically separated by distances
that are large compared to their width in position space,
$\ell_{\rm hop}\gg \ell_{\rm wp}$. Thus, we may find the positions
of several low-energy states
by using the following strategy.
\begin{itemize}
\item[(i)] For fixed parameter set $\vecalpha^{(1)}\equiv \veca$
  define the global low-pass filter
  \begin{equation}
    \Gamma_{\veca}(\vec{r}\,{}^{\prime}-\vec{r}\,)=
    \left|\varphi_{\veca}(\vec{r}\,{}^{\prime}-\vec{r}\,)\right|^2  \; .
  \end{equation}
  A first estimate for the particle positions is given by the minima of
  \begin{equation}
    W_{\rm R}^{\varphi}(\vec{r},\veca)=
    \int {\rm d}^3r'     \Gamma_{\veca}(\vec{r}\,{}^{\prime}-\vec{r}\,)
    {\cal V}_{\rm R}(\vec{r}\,{}^{\prime}) \; ,
  \end{equation}
  see eq.~(\ref{eq:convolution3dvariational}).
  The resulting minima are denoted by $\vec{s}_j^{\,(1)}$, $j=1,2,\ldots\;$.
\item[(ii)] Iterate to convergence the following steps:
  For given $\vec{s}_j^{\,(n)}$, optimize the variational parameters
  to find a new set $\vecalpha^{(n+1)}$; for this new set of parameters,
  optimize the position to obtain $\vec{s}_j^{\,(n+1)}$.
\end{itemize}
As a result, we obtain the variational landscape in the vicinity
of the minima $\vec{s}_j^{\;{\rm opt}}$ as
\begin{equation}
  W_{\rm R}(\vec{r},\vecalpha^{\rm opt})
  =
  \int {\rm d}^3r' \left|\varphi_{\vecalpha^{\rm opt}}(\vec{r}-\vec{r}\,{}^{\prime})\right|^2
    {\cal V}_{\rm R}(\vec{r}\,{}^{\prime})
  \label{eq:convolution3dvariationalmany}
\end{equation}
in the regions $|\vec{r}-\vec{s}_j^{\;{\rm opt}}|\lesssim \ell_{\rm wp}$.

\subsubsection{Gauss filter}
\label{subsubsec:variational-theory-Gauss}

To illustrate the concept of a local low-pass filter, we apply a Gauss filter
to the one-dimensional localization problem,
\begin{equation}
  \varphi(x'-x,\omega)
  = \left(\frac{m\omega}{\pi\hbar}\right)^{1/4}
  \exp\left(-\frac{m\omega(x'-x)^2}{2\hbar}\right)\;,
\label{eq:variationalphi}
\end{equation}
where $x'$ is the position of the wave packet and $\omega$
characterizes its width
$\ell_{\rm G}=\sqrt{\hbar^2/(2m \hbar\omega)}$,
so that we must optimize only one internal parameter,
$\alpha_1\equiv \omega$.
The variational energy reads
\begin{equation}
  E_{\rm R}^{\rm var}(x',\omega)=
\frac{\hbar\omega}{4}
  + \int_{-\infty}^{\infty} \frac{{\rm d} k}{2\pi} e^{{\rm i}k x'}
  \exp\left(-\frac{(k \ell_{\rm G})^2}{2}\right) \tilde{\cal V}_{\rm R}(k)
  \;,
  \label{eq:Evarinmomentumspace}
\end{equation}
using Fourier transformation, where we used
that the variational kinetic energy $t_{\rm G}= \hbar \omega/4$
for a Gaussian wave function.

The integral in eq.~(\ref{eq:Evarinmomentumspace})
can be calculated very efficiently by Fast-Fourier-Transformation so that
the minima of the variational energy can be found
from a scan of the two-dimensional
$(x',\omega)$ space.
In Fig.~\ref{fig:Gaussianlocalfilter} we show
the results for the first ten localized states.
It is seen that the positions of the lowest-lying states are very well
reproduced, and that the variational energies provide
a reasonable estimate for the exact values.

The energy estimate can be improved systematically by using a sum of
several Gaussians with different widths as variational state.
We shall not pursue this approach here.

\subsubsection{Gauss-Lorentz filter}

The effective potential deviates from its quadratic shape when
$|x-x'|\gtrsim \ell_{\rm G}$. It approaches the shape of a potential barrier
with constant height so
that the wave function decays exponentially, as is also seen in reciprocal space
in eq.~(\ref{eq:conditionsonGammatildek}).
Therefore, it is preferable to combine a Gaussian and a Lorentzian filter
so that
the variational energy is determined from the minima of the
function
\begin{eqnarray}
  E_{\rm R}^{\rm var}&=& t_{\rm GL} + v_{\rm GL} \; ,\nonumber \\
  v_{\rm GL} &=&
\int_{-\infty}^{\infty} \frac{{\rm d} k}{2\pi}
\frac{e^{{\rm i}k x'}}{1+\eta (k\ell_{\rm G})^2}
e^{-\frac{(1-\eta)(k\ell_{\rm G})^2}{2}} \tilde{\cal V}_{\rm R}(k)
  \;,\nonumber \\
  \varphi(x) &=& \left[\int_{-\infty}^{\infty} \frac{{\rm d} k}{2\pi}
    \frac{e^{{\rm i}k x}}{1+\eta (k\ell_{\rm G})^2}
    e^{-\frac{(1-\eta)(k\ell_{\rm G})^2}{2}}\right]^{1/2}
  \label{eq:EvarinmomentumspaceGausLorentz}\\
    t_{\rm GL} &=& \frac{\hbar^2}{2m}
    \int_{-\infty}^{\infty} {\rm d} x \left[\varphi'(x)\right]^2 \nonumber \; ,
     \end{eqnarray}
where we abbreviated $E_{\rm R}^{\rm var}(x',\omega,\eta)\equiv   E_{\rm R}^{\rm var}$.
The optimal values for $\eta_j$ for the localized states at $x_j'$
are found to lie in the range $0.6\leq \eta_j\leq 0.8$.

\begin{figure}[t]
\includegraphics[width=\linewidth]{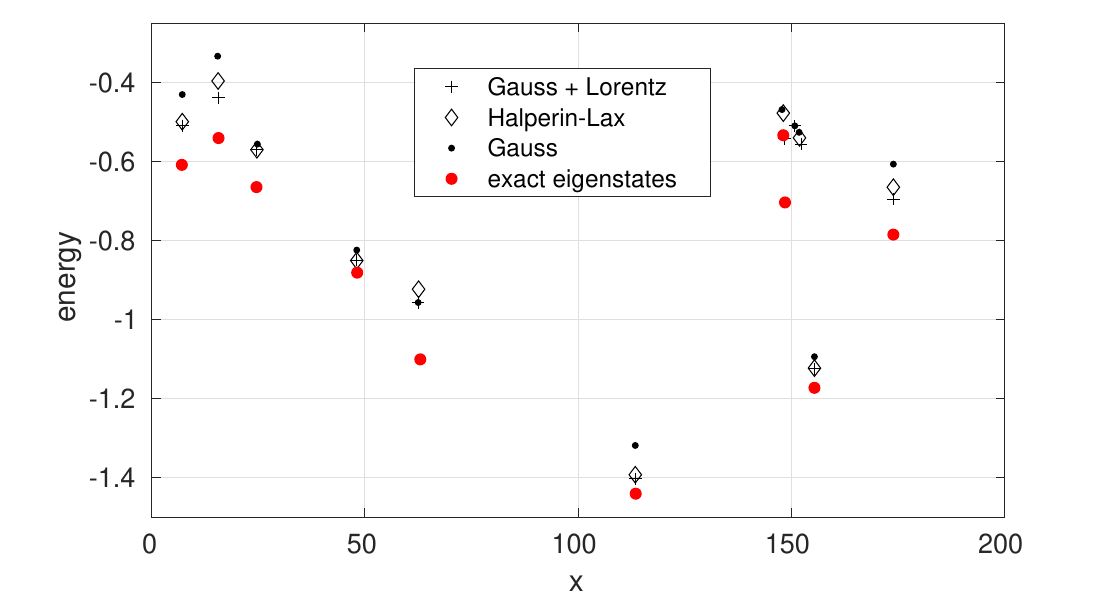}
\caption{Positions of the ten lowest-lying localized states and their energies
  (red points) for the random potential
  of Fig.~\protect\ref{fig:potentialandstates}.
  The black dots, crosses, and diamonds
  indicate the positions and energies for Gauss, Gauss-Lorentz, and
  Halperin-Lax filters, respectively.\label{fig:Gaussianlocalfilter}}
\end{figure}

In Fig.~\ref{fig:Gaussianlocalfilter} we also include the results
from the Gauss-Lorentz
filter. Since we enlarge the variational space, the energies from
the Gauss-Lorentz states are closer to the exact results.
Note that a combination of Gaussian and Lorentzian variational
states is similar in spirit
to the application of the LLT to a random potential with Gaussian averaging,
see eq.~(\ref{eq:random_potential_correlated}).

\subsubsection{Halperin-Lax filter}
\label{subsubsec:variational-theory-Halperin-Lax}

Halperin and Lax used a variational approach to calculate the density of states
deep in the band tails~\cite{Halperin1966}.
Their approach provides the functional form that best fits
the typical low-energy localized state. In three dimensions,
the Halperin-Lax state~$f(\vec{r}\,)$
is obtained from
the solution of the third-order equation
\begin{equation}
  T^{\rm op}(\vec{r}\,) f(\vec{r}\,) -\mu  \frac{{\cal V}^2(\ell_{\rm pf})^3}{12}
  [f(\vec{r}\,)]^3= E f(\vec{r}\,)
\label{eq:Lax-Halperin equation}
\end{equation}
for the case of Gaussian white noise.
This equation is also obtained from the method of optimal fluctuations
by Baranovskii and Efros~\cite{Baranovskii1978}.

We introduce the length unit $\ell_{\rm wp}$, $\vec{r}=\ell_{\rm wp}\vec{x}$,
and the energy unit ${\cal T}$ so that $e=E/{\cal T}$ and $\bar{\cal V}
= {\cal V}/{\cal T}$. Moreover,
we set $\mu= \tilde{\mu}/{\cal T}$,
and $f(\vec{r}\,)=F(\vec{x}\,)/\sqrt{\ell_{\rm pf}^3}$.
Then, the dimension-less Halperin-Lax equation reads
\begin{equation}
  -\sum_{i= 1}^3 \frac{\partial ^2 F(\vec{x}\,)}{\partial x_i^2}
  -\tilde{\mu } S_L [F(\vec{x}\,)]^3= e F(\vec{x}\,)  \; ,
\label{eq:Lax-Halperin equation-no-dimension}
\end{equation}
where
\begin{equation}
S_L= \frac{\bar{\cal V}^2}{12}\left(\frac{\ell_{\rm pf}}{\ell_{\rm wp}}\right)^3
\end{equation}
is the dimensionless impurity interaction strength in three dimensions.

In one dimension, the dimensionless Halperin-Lax equation reads
\begin{equation}
  -F''(x) -\tilde{\mu } S_L [F(x)]^3= e F(x)
\label{eq:Lax-Halperin-equation-no-dimension-onedim}
\end{equation}
with the normalized solution~\cite{Halperin1966}
\begin{equation}
  F(x) = \sqrt{\frac{\kappa}{2}} \frac{1}{\cosh(\kappa x)}\; ,
\label{eq:Halperin-Lax-state}
\end{equation}
where the decay length $\ell_{\rm HL}= 1/\kappa$
depends on the energy~$e$,
\begin{equation}
  \kappa\equiv \kappa(e)= \sqrt{-e} = \frac{1}{\ell_{\rm HL}}
  \label{eq:kappofe}
\end{equation}
and $\tilde{\mu}=4\kappa/S_L$.

For low energies,
the resulting variational
density of states $\rho_{\rm var}(E)$
in one dimension agrees in its functional form
with the exact solution~\cite{exactDOSreference,PhysRev.139.A104,LaxRevModPhys}
up to a factor $1/\sqrt{5}$~\cite{Halperin1966}.
Extensions to variational states that include perturbative corrections
to the form~(\ref{eq:Halperin-Lax-state}) permit to obtain
the exact
prefactor to the low-energy
density of states~\cite{Halperin1967,PhysRev.148.741}.

Apparently, the Halperin-Lax wave function~$F(x)$
in eq.~(\ref{eq:Halperin-Lax-state})
interpolates between Gaussian behavior for small distances and
exponential behavior
for large distances using only a single parameter.
The results for the Halperin-Lax filter where we replace $\varphi(x)$
in eqs.~(\ref{eq:RitzVP})--(\ref{eq:calVvariational}) by $F(x)$
are also included in Fig.~\ref{fig:Gaussianlocalfilter}.
They are better than those obtained from
the Gauss filter and comparably good as those obtained
from the Gauss-Lorentz filter.

The Halperin-Lax filter offers the following advantages.
First, it does not find too many states, i.e.,
as seen in Fig.~\ref{fig:Gaussianlocalfilter}, there is one Halperin-Lax state
for every exact localized state whereas Gauss and Gauss-Lorentz
find an additional state
in the region of $x\approx 150$. We observed this behavior also
for other realizations~${\rm R}$.
Second, the Halperin-Lax wave function gives good variational energies
using only a single variational parameter for its decay length,
$\ell_{{\rm HL},j}= 1/\kappa_j$ for the $j$th minimum.
Third, the Halperin-Lax wave function provides the
optimal form for localized low-energy states in a statistical sense.
Therefore, we come to the conclusion that the Halperin-Lax filter is the most
appropriate one-parameter local low-pass filter.

\subsection{Halperin-Lax global low-pass filter}
\label{subsec:globalLPF}

A global low-pass filter includes a finite set of length scales that
are independent
of the realization~${\rm R}$. To determine their values, we
should optimize
the agreement with physical quantities that are obtained by averaging
over many realizations, e.g.,
the density of states $\rho(E)$ in eq.~(\ref{eq:dosdef}).
Below we shall focus on the construction of a global
Halperin-Lax filter in one dimension.

\subsubsection{Motivation}

\begin{figure}[t]
\includegraphics[width=\linewidth]{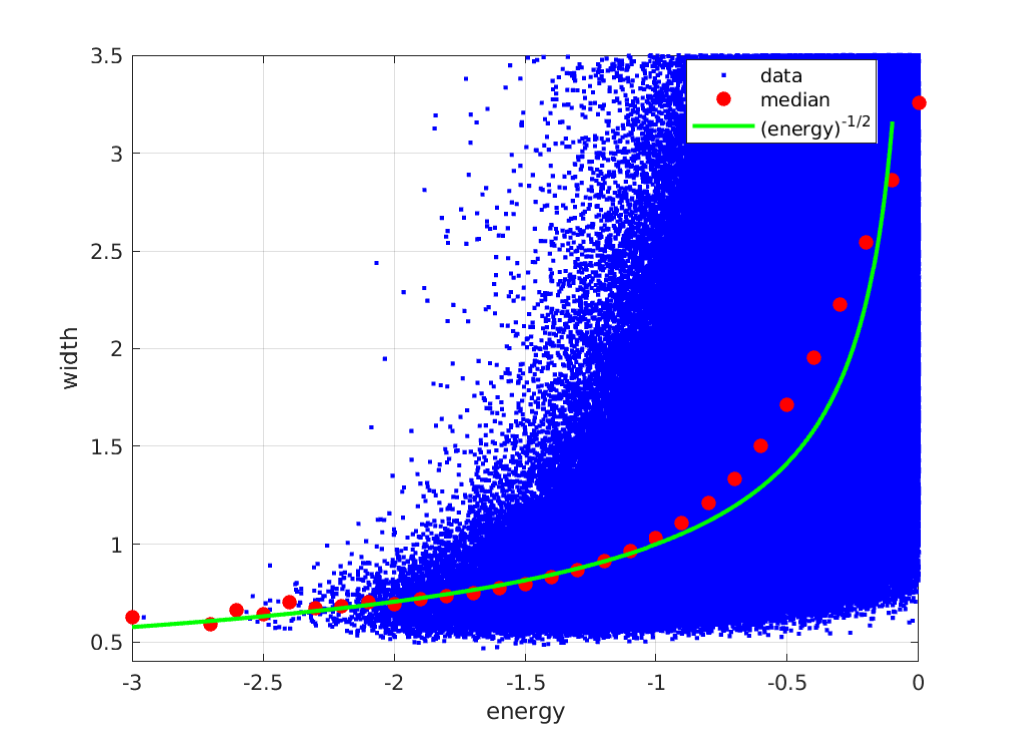}
\caption{Relation between the exact energies for the
  lowest-lying localized states $e_j$ and the corresponding
  standard deviations $\ell_{{\rm HL},j}$ of the wave packets in the
  Halperin-Lax wave function in one dimension,
  see eq.~(\protect\ref{eq:Halperin-Lax-state}).
  Also shown is the median of the data (red dots) and the Halperin-Lax
  relation $\ell_{\rm HL}(e)=1/\sqrt{-e}$,
  see eq.~(\protect\ref{eq:kappofe}).\label{fig:kappe-e-relation}}
\end{figure}

Eq.~(\ref{eq:kappofe}) suggests that the width of the wave function
increases as a function of energy. As expected,
the lowest-lying states have the smallest spread in position space.
For a given realization~${\rm R}$,
the surrounding of the potential well also plays an important role.
Therefore, the relation between between binding energy and wave-function width
is not always so simple as suggested in eq.~(\ref{eq:kappofe}).
This is shown in Fig.~\ref{fig:kappe-e-relation},
where we display the relation between
the lowest exact energies $e_{l}$ and the corresponding
standard deviations $\ell_{{\rm HL},j}$ of the wave packets
for $\ell_{\rm HL}$ for 26.000 realizations~${\rm R}$ of systems
with $L_s=2.000$ and $L=200$ for $S_L=1$.

As seen from Fig.~\ref{fig:kappe-e-relation}, the median follows the
relation~(\ref{eq:kappofe}) but there is considerable spread in the values
$\ell_{\rm HL}$. Nevertheless, since the typical energy values are of the
order $S_L^{2/3}{\cal T}$, the full variety of local
variational wave functions can be approximated by a prototypical wave function
when $\ell_{\rm HL}/(\ell_{\rm wp}S_L^{1/3})$
lies in the fairly small interval $I_{\ell_{\rm HL}}= [0.8,1.2]$.
Therefore, it is permissible to replace
the set of local Halperin-Lax low-pass filters
by a global Halperin-Lax low-pass filter,
\begin{equation}
  \Gamma^{\rm HL}(y)
  =\frac{1}{2 \ell_{\rm  HL}} \frac{1}{\cosh^2(y/\ell_{\rm HL})} \; .
  \label{eq:globalHLfilter}
    \end{equation}
The corresponding potential landscape becomes
\begin{equation}
  W_{\rm R}^{\rm HL} (x) = \int {\rm d} x'  \Gamma^{\rm HL}(x-x') {\cal V}_{\rm R}(x')
   \label{eq:globalHLpotential}
\end{equation}
for a given realization~${\rm R}$ of the random white-noise potential.
The remaining task is to determine the `best' value for $\ell_{\rm HL}$.

\subsubsection{Tail optimization}

One strategy suggested by Fig.~\ref{fig:kappe-e-relation} is to optimize
the states in the low-energy tail. To optimize the tail states,
one should choose
\begin{equation}
  \frac{\ell_{\rm HL}^{\rm tail}S_L^{1/3}}{  \ell_{\rm wp}}=0.8
  \label{eq:globalellHL}
  \end{equation}
in eq.~(\ref{eq:globalHLfilter}).
Since there are exponentially few states in the tails, this is not
the best strategy
to describe the system at elevated temperatures. Therefore,
we shall not elaborate this choice any further.

\subsubsection{Classical state counting}

\begin{figure}[t]
  \includegraphics[width=\linewidth]{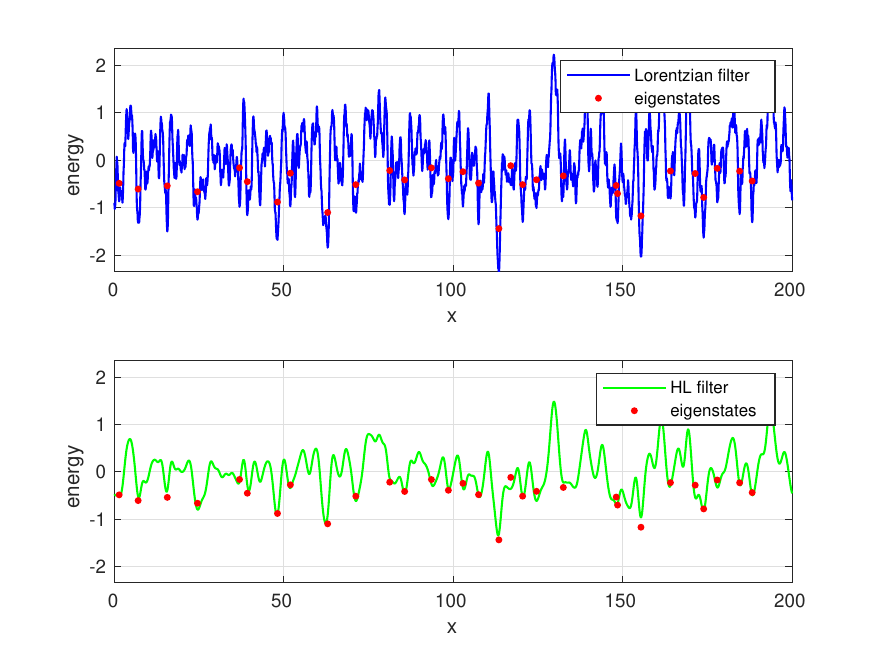}
\caption{(a) Effective potential $W_{\rm R}^{\rm L}(x)$ for a global
  Lorentzian low-pass filter for $\ell_{\rm L}=0.436\,\ell_{\rm wp}$,
  see eq.~(\protect\ref{eq:convolutionLorentz})
  and Fig.~\protect\ref{fig:Lorentzian}.
  (b) Effective potential $W_{\rm R}^{\rm HL}(x)$ for a global
  Halperin-Lax low-pass filter for $\ell_{\rm HL}=1.2\, \ell_{\rm wp}$ ($S_L=1$),
  see eq.~(\protect\ref{eq:globalHLpotential}).
  Also shown in both figures are the exact positions and energies
  of the 30~states lowest in energy.\label{fig:globalHLpotential}}
\end{figure}

In Fig.~\ref{fig:globalHLpotential} (upper part) we reproduce
the Lorentzian landscape from LLT with $\ell_{\rm L}=0.436\,\ell_{\rm wp}$
for the realization of the random potential shown
in Fig.~\ref{fig:potentialandstates}(a),
as already shown in Fig.~\ref{fig:Lorentzian}.
Here, we add the exact positions and energies of the 30~lowest-lying states.
It is seen that there is a reasonable overall agreement between the positions
and depth of the minima in the LLT landscape and the exact results.
However, the potential is still noisy and some minima do not support an
exact eigenstate.

The Halperin-Lax potential landscape shown in the lower part of
Fig.~\ref{fig:globalHLpotential}
is much smoother because $\ell_{\rm HL}=1.2\, \ell_{\rm wp}$ at $S_L= 1$ is almost
three times larger than $\ell_{\rm L}$. More importantly, all minima
for energies below $e=-0.5$ correspond to a localized state
in the exact solution of the Schr\"odinger equation.
If we choose $\ell_{\rm L}= 1.2\, \ell_{\rm wp}$ for the Lorentzian filter,
the corresponding effective potential becomes less noisy but there still
are quite a number of local minima that do not correspond
to an exact bound state. The reason for this behavior is the cusp
at $x= x'$ in the Lorentzian filter, see eq.~(\ref{eq:convolutionLorentz}).
The smooth Halperin-Lax filter in eq.~(\ref{eq:globalHLfilter})
avoids these artificial potential fluctuations.

As seen from Fig.~\ref{fig:globalHLpotential}, the Halperin-Lax filter not only
reproduces the spatial positions of the localized states but also their
energy with very good accuracy.
This indicates that a classical counting of states will
provide a reasonable estimate for the integrated particle density $N(e)$ below
the energy~$e=E/{\cal T}$
that is known exactly for random white-noise potential
in one dimension~\cite{exactDOSreference,PhysRev.139.A104,LaxRevModPhys},
\begin{eqnarray}
  N(e)&=&\int_{-\infty}^{e}{\rm d} \epsilon \rho(\epsilon) \; ,
  \label{eq:IntegratedexactDOSasymptotic}\\
  &=& \frac{1}{2^{1/3}\pi^2} \left[
[\Ai(-2^{2/3}e)]^2+ [\Bi(-2^{2/3}e)]^2
    \right]^{-1}, \nonumber
\end{eqnarray}
where $\Ai(x)$ and $\Bi(x)$ are the Airy functions of the first and
second kind,
\begin{eqnarray}
  \Ai (x)&=&{\frac {1}{\pi }}\int \limits _{0}^{\infty }{\rm d}t
  \cos \left({\frac {t^{3}}{3}}+xt\right) \; , \\
       \Bi (x)&=&{\frac {1}{\pi }}\int \limits _{0}^{\infty }\,{\rm  d}t
       \left[\exp \left(-{\frac {t^{3}}{3}}+xt\right)
       +\sin \left({\frac {t^{3}}{3}}+xt\right)\right] \; .\nonumber
\end{eqnarray}
We concisely re-derive the result~(\ref{eq:IntegratedexactDOSasymptotic})
in appendix~\ref{app:NofE}. Furthermore, in the same appendix~\ref{app:NofE} we derive the expression for the energy-dependent localization length $a(e)$ in the white-noise one-dimensional potential.

Treating $W_{\rm R}^{\rm HL}(x)$ as a classical potential,
we denote the minima at position $x_i$ with energy $e_i$
by the set $M_{\rm R}^{\rm HL}= \{(x_{i,{\rm R}},e_{i,{\rm R}}) \}$.
Then, the number of states up energy~$e$ is given by
classical counting
\begin{eqnarray}
  N_{\rm cl}^{\rm HL}(e)&=& \langle N_{\rm cl,R}(e)\rangle \;, \nonumber \\
  N_{\rm cl,R}(e) &= &\frac{1}{L} \sum_{M_{\rm R}^{\rm HL}} \Theta(e-e_{i,{\rm R}}) \;,
  \label{eq:countminimaforNofe}
\end{eqnarray}
where $\Theta(x)$ is the Heaviside step function.
$N_{\rm cl,R}(e)$
simply counts the minima with energy up to~$e$
in the realization~${\rm R}$ of the Halperin-Lax potential.

\begin{figure}[ht]
  \includegraphics[width=\linewidth]{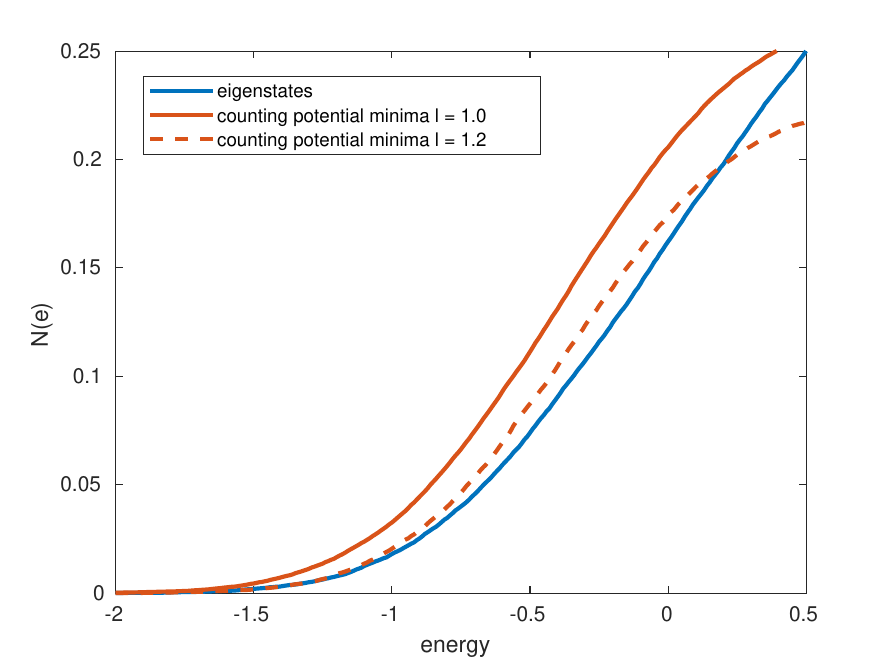}
  \caption{Exact result for the integrated particle density up to energy~$e$,
    $N(e)$ from eq.~(\protect\ref{eq:IntegratedexactDOSasymptotic}),
    in comparison with the classical
    counting of minima, $N_{\rm cl}(e)$ from eq.~(\ref{eq:countminimaforNofe}),
    for the Halperin-Lax potential for $\ell_{\rm HL}= 1.2\,\ell_{\rm wp}$
    (dashed line)
    and $\ell_{\rm HL}= 1.0\,\ell_{\rm wp}$ (full line)
    ($S_L= 1$). Averages are taken over 1.000
    realizations~${\rm R}$.\label{fig:exactvsclasicalcounting}}
\end{figure}

The exact result $N(e)$ in one
dimension~(\ref{eq:IntegratedexactDOSasymptotic})
and the result from classical counting of minima~(\ref{eq:countminimaforNofe})
are compared in Fig.~\ref{fig:exactvsclasicalcounting}.
It is seen that the classical counting works very well
deep in the tail, $e \lesssim -1$, and still gives reasonable results
up to the band edge, $e_{\rm be}= 0$.
The results are best for $\ell_{\rm HL}=1.2\, \ell_{\rm wp}$
and slightly poorer when we choose $\ell_{\rm HL}=\ell_{\rm wp}$.

The success of the classical counting of minima shows that
the Halperin-Lax potential ought to be viewed as a suitable effective potential
for {\em classical\/} particles.

\subsubsection{Semi-classical state counting}

In Ref.~[\onlinecite{Arnold2016}], Arnold et al.\ proposed to
use the LLT potential in the semi-classical Weyl expression~\cite{Weyl1912}
to approximate the integrated particle density $N(e)$.
For particles in a potential landscape $W_{\rm R}(x)$
in one dimension it reads~\cite{Arnold2016}
\begin{eqnarray}
  N_{\rm W}(E)&=&\frac{1}{L}
    \frac{\sqrt{2m}}{\pi\hbar}
  \langle \int {\rm d} r \sqrt{E-W_{\rm R}(r)}
  \Theta\left(E-W_{\rm R}(r)\right) \rangle \; , \nonumber \\
  N_{\rm W}(e)&=&\frac{1}{L}\frac{1}{\pi} \langle
  \int {\rm d} x \sqrt{e-\bar{W}_{\rm R}(x)}
  \Theta\left(e-\bar{W}_{\rm R}(x)\right) \rangle \; ,
  \label{eq:NofEfromWeyl}
\end{eqnarray}
where the second line gives
the dimensionless expression where energies are measured
in units of ${\cal T}$.

\begin{figure}[b]
\includegraphics[width=\linewidth]{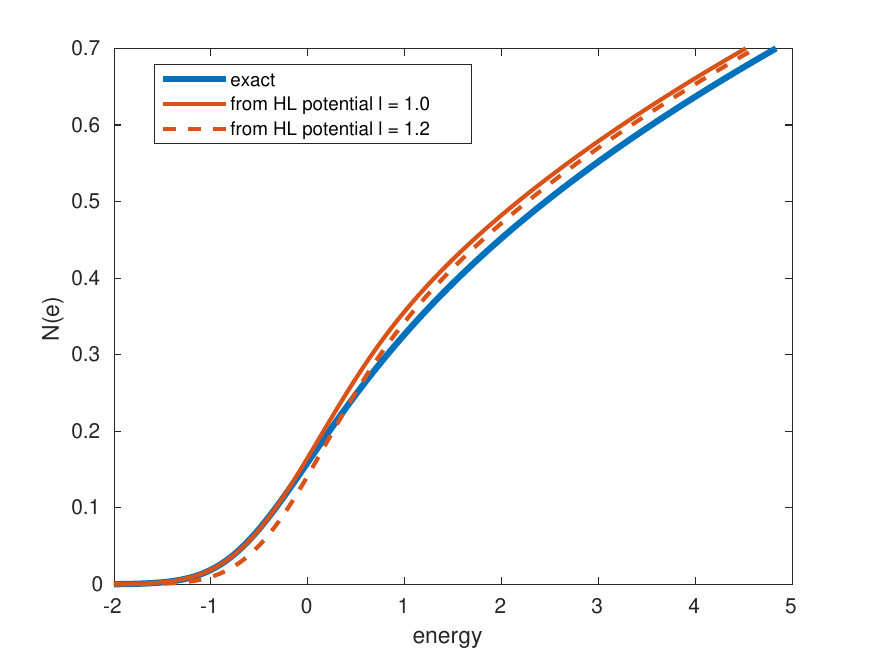}
\caption{Integrated particle density $N(e)$ up to energy~$e=E/{\cal  T}$
  in a one-dimensional
  random white-noise potential.
  The exact result~(\ref{eq:IntegratedexactDOSasymptotic})
  is compared with the semi-classical state
  counting using eqs.~(\protect\ref{eq:Wbarshifted})
  and~(\protect\ref{eq:NofEfromWeylshifted})
  for a Halperin-Lax filter with $\ell_{\rm HL}= 1.2\,\ell_{\rm wp}$
  (dashed line) and
with $\ell_{\rm HL}= 1.0\,\ell_{\rm wp}$ (full line) ($S_L= 1$),
  averaged over 1.000 configurations with 2.000 sites
  for $L= 200$.\label{fig:integratedDOS}}
\end{figure}

Obviously, one should not use $W_{\rm R}^{\rm HL}(x)$
in eq.~(\ref{eq:NofEfromWeyl}) because the Weyl formula starts from the
assumption that the total energy is the sum of the
kinetic energy and the potential energy.
For given $\ell_{\rm HL}$, the kinetic energy of the variational state
is a constant,
\begin{equation}
  \bar{T}_0 = \frac{1}{{\cal T}}\frac{\hbar^2}{2m}
\int {\rm d} r \left[F'(r)\right]^2
  =
  \frac{1}{3} \left(\frac{\ell_{\rm wp}}{\ell_{\rm HL}}\right)^2
\end{equation}
in units of ${\cal T}$.
Therefore, we must shift the classical potential $\bar{W}_{\rm R}(x)$ by
$\bar{T}_0$,
\begin{equation}
\widetilde{W}_{\rm R}(x) =
\bar{W}_{\rm R}(x)-\bar{T}_0 \; ,
\label{eq:Wbarshifted}
\end{equation}
and we employ
\begin{equation}
  N_{\rm scl}(e)= \frac{1}{L}
\frac{1}{\pi} \langle
  \int {\rm d} x \sqrt{e-\widetilde{W}_{\rm R}(x)}
  \Theta\left(e-\widetilde{W}_{\rm R}(x)\right) \rangle
  \label{eq:NofEfromWeylshifted}
\end{equation}
for an appropriate semi-classical approximation of the number of states
up to energy~$e$.

In Fig.~\ref{fig:integratedDOS}
we compare the exact result for the integrated particle density up to energy~$e$,
eq.~(\ref{eq:IntegratedexactDOSasymptotic}),
with the (shifted) semi-classical expression~(\ref{eq:NofEfromWeylshifted})
for the Halperin-Lax potential at $\ell_{\rm HL}= 1.2\, \ell_{\rm wp}$
and $\ell_{\rm HL}= 1.0\,\ell_{\rm wp}$,
obtained from an average over 1.000
configurations for $L_{\rm s}= 2.000$ sites for system length $L= 200$ ($S_L= 1$).
The comparison shows that the semi-classical approximation
faithfully reproduces the exact result up to the band edge $e_{\rm be}=0$,
and remains
reasonable even beyond. The results for $\ell_{\rm HL}= 1.2\,\ell_{\rm wp}$ are
slightly poorer than those for $\ell_{\rm HL}=1.0\,\ell_{\rm wp}$
but still astonishingly good.
Thus, the shifted Halperin-Lax potential provides a sound basis
for a semi-classical description of particles in a localization landscape.

\section{Conclusions}
\label{sec:conclusions}

In this work, we discussed a number of approaches to
deeply localized states in disordered systems.
Therefore, after we put our work into perspective,
we summarize the proposed strategies. We close our presentation
with a short outlook.

\subsection{Perspective}

In disordered systems, the potential experienced by the carriers
fluctuates on an atomic length scale, $\ell_{\rm pf}= 1\, \hbox{\AA}$.
In contrast, the localized low-energy states
spread over a much larger length scale, namely
$\ell_{\rm wp}\approx 10\ell_{\rm pf}\gg \ell_{\rm pf}$.
In addition, states with comparable energy are well separated in space,
$\ell_{\rm hop}\gg \ell_{\rm wp}$.
Therefore, it is advisable to take advantage
of the separation of length and energy scales for localized states
in the band tails.

When charge carriers move between these localized states,
the microscopic length scale $\ell_{\rm pf}$ becomes irrelevant, and only
a coarse-grained potential landscape remains to be considered.
It is the goal of the localization landscape
theory (LLT)~\cite{Arnold2016,LL1_2017,LL2_2017,LL3_2017},
in its application to electronic states in disordered media,
to derive systematically useful landscape
$W_{\rm R}(\vec{r}\,)$ for
a given realization~${\rm R}$ of the fluctuating potential
${\cal V}_{\rm R}(\vec{r}\,)$. Landscapes are useful if they
reliably reproduce the
spatial positions and energies of the low-lying localized states.

The LLT application to the Schr\"odinger equation
is based on the solution of eq.~(\ref{eq:U}).
The corresponding LLT potentials
have been applied successfully to a variety of disordered systems.
In this work, we first showed that the LLT corresponds to a global Lorentzian
low-pass filter where the solution of eq.~(\ref{eq:U}) fixes
the single length-scale $\ell_{\rm L}$.
This observation cleared the way to investigate the full range of
decay lengths and a broader class of low-pass filters
to extend and improve the LLT{-inspired methods} systematically.
The resulting strategies are summarized as follows.

\subsection{Strategies}

For low temperatures, the low-energy states deep in the band tails
are of particular importance. There are several ways to
determine their physical properties, each of which having its merits and
disadvantages.

\begin{itemize}
\item[(i)] Exact results are obtained from the solution of the
  Schr\"odinger equation. From there, expectation values for physical quantities
  are calculated exactly for a given configuration of the disorder potential,
  and averages over many realizations are required.

  This approach is very time-consuming and not feasible
  in three dimensions, especially when the calculations
  are part of a self-consistent loop, i.e., when they need to be repeated
  a large number of times.
\item[(ii)] The localization landscape theory (LLT) provides
  a far less time-consuming
  alternative~\cite{Arnold2016,LL1_2017,LL2_2017,LL3_2017}
  because it provides a smooth potential $W_{\rm R}(\vec{r}\,)$
  that can be used for (semi-)classical considerations.

  However, the LLT still requires the solution of eq.~(\ref{eq:U})
  for each realization. Moreover, the LLT foundations and limits for its
  applicability remained unclear.
\item[(iii)] The structure of the low-energy wave functions of the Hamiltonian
justifies the application of a low-pass filter to the potential.
{In particular, application of the LLT} corresponds to a
Lorentzian low-pass filter, see eq.~(\ref{eq:convolutionLorentz})
for the definition of the landscape $W_{\rm R}^{\rm L}(\vec{r}\,)$.

The value for the decay length $\ell_{\rm L}$ in the Lorentzian low-pass
filter can be estimated from the solution
  of the LLT eq.~(\ref{eq:U}) for a single realization.
  Alternatively, one may take $\ell_{\rm L}$ as the width of the
  ground-state wave packet of the Hamiltonian for some configuration~${\rm R}$,
  or simply set $\ell_{\rm L}=\ell_{\rm wp}$. For an expression of $\ell_{\rm wp}$
  as a function of the system parameters, see appendix~\ref{app:B}.
  In this way, the solution of eq.~(\ref{eq:U}) becomes superfluous.
\item[(iv)]  When individual localized states
  are required for a given realization of the random potential,
  local low-pass filters (Gauss, Gauss-Lorentz, Halperin-Lax)
  can be applied.
  These methods are very efficient because
  expectation values can be calculated using Fast-Fourier-Transformation
  techniques, and the optimized states provide a fairly accurate
  variational description of deeply localized states.
\item[(v)] The best global low-pass filter is not provided by
  the Lorentzian filter. The optimal filter in a statistical sense
  is obtained from the solution of the Halperin-Lax
  equation~(\ref{eq:Lax-Halperin equation});
  the latter equation is not limited to random white-noise potentials.

  Using the Halperin-Lax low-pass filter as global filter
  leads to the Halperin-Lax potential landscape
  $W_{\rm R}^{\rm HL}(\vec{r}\,)$, see eqs.~(\ref{eq:globalHLfilter})
  and~(\ref{eq:globalHLpotential}) for one spatial dimension
  and a random white-noise potential.
  The Halperin-Lax potential can be viewed as potential
  landscape for classical particles. A classical state counting
  reproduces the exact integrated particle density $N(e)$  at low energies
  surprisingly well.
\item[(vi)] When the kinetic energy is taken into account,
  the shifted Halperin-Lax potential
  $W_{\rm R}^{\rm HL}(\vec{r}\,)-T_0$, see eq.~(\ref{eq:Wbarshifted}),
  can be used for a semi-classical
  description of particles in a random potential.
  The Weyl approximation for the density of states using
  $W_{\rm R}^{\rm HL}(\vec{r}\,)-T_0$ reproduces
  the exact result for the integrated particle density $N(e)$
  in one dimension with very good accuracy.
\end{itemize}

\subsection{Outlook}

In this work, we derived the concept of a global
Halperin-Lax low-pass filter
for the localization landscape theory,
and successfully applied it to particles in a one-dimensional
random white-noise potential.
It remains to be tested that
the size of a typical wave packet can be used as characteristic length
for the Halperin-Lax filter also in two and three dimensions.

The Halperin-Lax filter can be found from
the solution of a integro-differential equation that reduces
to a non-linear differential equation for the case of white noise.
Therefore, the localization landscape obtained with a Halperin-Lax filter
can readily be applied to
random potentials with correlations.
In this case, more than one length scale $\ell_{\rm wp}$
might be necessary to parameterize the low-pass filter appropriately.

For transport, not only the typical distance $\ell_{\rm hop}$ between
energetically low-lying states is important but also the
typical potential landscape between them determines the tunnel
probabilities between those states.
Further studies are necessary to clarify whether or not the
localization landscape with Halperin-Lax filter is useful to describe
hopping transport at finite temperatures.

The main focus of the LLT in solids is not the single-particle density of states
but rather the particle density $n(\vec{r},T)$ that depends on space~$\vec{r}$
and temperature~$T$.
In our subsequent paper~\cite{Nenashev_Paper_2}
we calculate $n(\vec{r},T)$ without solving the
Schr\"odinger equation. Thereby, the full quantum-mechanical problem
is reduced to the quasi-classical description of $n(\vec{r},T)$
in a temperature-dependent effective potential $W(\vec{r}\,,T)$.
Our approximate results
for $n(\vec{r},T)$ and for the carrier mobility at elevated temperatures favorably
compare with those from
the exact solution of the Schr\"odinger equation~\cite{Nenashev_Paper_2}.
The temperature-dependence of $W(\vec{r}\,,T)$ leads to superior results
in comparison with the temperature-independent LLT landscape $W(\vec{r})$.

\begin{acknowledgments}
A.N. thanks the Faculty of Physics of the Philipps Universit\"at Marburg
  for the kind hospitality during his research stay. S.D.B. and K.M. acknowledge financial support by the Deutsche
Forschungsgemeinschaft (Research Training Group ``TIDE'', RTG2591)
as well as by the key profile area ``Quantum Matter and Materials
(QM2)'' at the University of Cologne. K.M. further acknowledges support
by the DFG through the project ASTRAL (ME1246-42).
\end{acknowledgments}

\appendix

\section{Random white-noise potential in one dimension}
\label{app:NofE}

\begin{figure}[b]
\includegraphics[width=\linewidth]{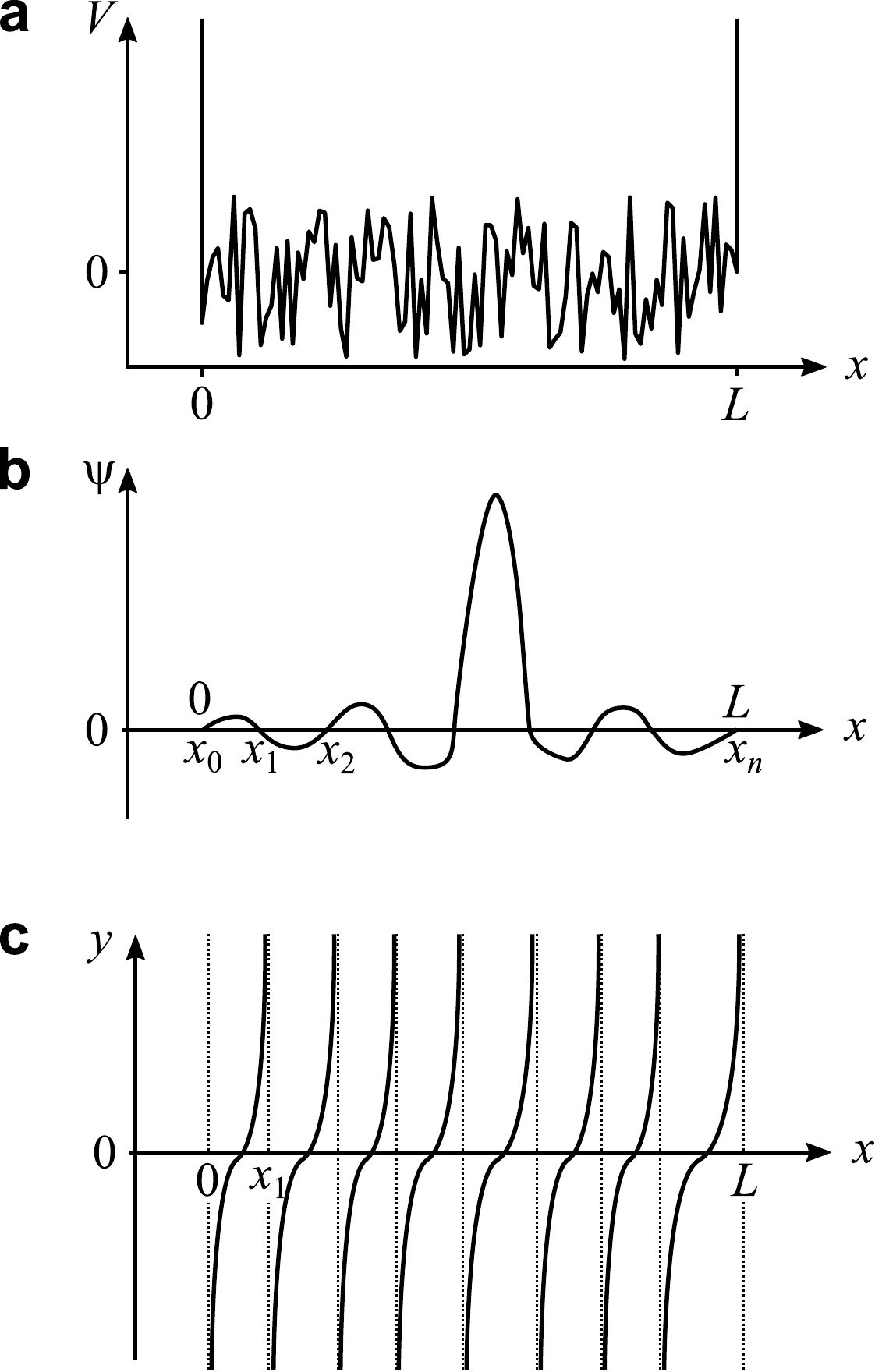}
\caption{Localized state in a one-dimensional random white-noise potential.
  (a) sketch of potential $V(x)$;
  (b) wave function $\psi(x)$ that corresponds to the $n$th eigenstate
  whose energy is approximately equal to $e$;
  (c) function $y(x) = - \psi'(x)/\psi(x)$.
  The particle is confined between infinite walls at $x=0$ and $x=L$.
  The zeros of the wave function are labeled by $x_1$, $x_2$,
  et cetera.\label{fig:dos_1d}}
\end{figure}

\subsection{Density of states}

The one-dimensional Hamiltonian eq.~(\ref{eq:defHreduzed})
gives rise to the Schr\"odinger equation
\begin{equation}
  	-\psi''(x) + V(x)\psi(x) = e \psi(x)\;,
\label{eq:app-SE}
\end{equation}
where $V(x)$ is a normalized white-noise potential,
\begin{equation}
	\langle V(x) \rangle = 0 \;, \quad
	\langle V(x_1) V(x_2) \rangle = \delta(x_1-x_2) \;.
        \label{eq:app-V-statistics}
\end{equation}
We consider a particle confined within a range $x\in[0,L]$, see
Fig.~\ref{fig:dos_1d}(a), implying open boundary conditions
\begin{equation}
  \psi(0) = \psi(L) = 0 \;.
  \label{eq:app-BC}
\end{equation}
In the limit $L\to\infty$, the energy spectrum of the particle becomes continuous.
Therefore, for a very large length $L$, any given energy $e$
can be approximated by some energy level in this system, i.e.,
some solution of the Schr\"odinger equation~(\ref{eq:app-SE}).
Let $n(e)$ denote the number of states up to this energy level,
counting from below, so that
\begin{equation}
  N(e) = \frac{n(e)}{L} \; .
  \label{eq:app-N-def}
\end{equation}
The aim of this Appendix is to derive
the expression~(\ref{eq:IntegratedexactDOSasymptotic})
for $N(e)$ in the limit of $L\to\infty$.

According to the oscillatory theorem, the wave function $\psi(x)$
that corresponds to the $n$th energy level has $n+1$ zeros,
$x_0 = 0$, $x_1$, $\ldots$, $x_n=L$, see Fig.~\ref{fig:dos_1d}(b).
Let us consider the function
\begin{equation} \label{eq:app-y-def}
	y(x) = -\frac{\psi'(x)}{\psi(x)} \; .
\end{equation}
A schematic plot of this function is shown in Fig.~\ref{fig:dos_1d}(c).

Taking the derivative $y'(x)$ in eq.~(\ref{eq:app-y-def})
and expressing $\psi''$ via eq.~(\ref{eq:app-SE}),
one can obtain a first-order differential equation for the function $y(x)$,
\begin{equation} \label{eq:app-y-equation}
	y'(x) = y(x)^2 + e - V(x) \; .
\end{equation}
For a further analysis of eq.~(\ref{eq:app-y-equation}),
it is convenient to interpret the variable $x$ as `time'.
Then, the function $y(x)$ can be understood as describing the movement
of some point along the `coordinate' $y$ in `time' $x$.
According to eq.~(\ref{eq:app-y-equation}), this movement is a combination
of drift with velocity
\begin{equation}
  v(y) = y^2 + e
  \label{eq:app-v-def}
\end{equation}
and a diffusion due to a Langevin force introduced by the white-noise term
in eq.~(\ref{eq:app-y-equation})~\cite{vanKampen}.

We determine the diffusion coefficient $D$
in the corresponding Fokker--Planck equation from the statistics of the
random white-noise potential $V(x)$.
To this end, we ignore the drift term,
and consider the diffusive equation of motion
with only the white-noise term present, $y' = -V(x)$.
Its solution is $y(x) = y(0) - \int_0^x V(x_1)\,{\rm d}x_1$. Therefore,
the mean square of the displacement along $y$ during the `time' $x$ is
\begin{eqnarray}
\left\langle [y(x)-y(0)]^2 \right\rangle &=&
\left\langle \int_0^x V(x_1)\,{\rm d}x_1 \int_0^x V(x_2)\,{\rm d}x_2 \right\rangle
\nonumber \\
&=& \int_0^x {\rm d}x_1 \int_0^x {\rm d}x_2  \left \langle V(x_1) V(x_2)
\right\rangle \nonumber \\
&=& x \; ,
\label{eq:app-mean-sq-y-1}
\end{eqnarray}
where we used the white-noise correlation function~(\ref{eq:app-V-statistics}).
By definition, the left-hand side of eq.~(\ref{eq:app-mean-sq-y-1})
is equal to $2Dx$~\cite{vanKampen}.
Hence, $2Dx = x$, i.e.,
\begin{equation} \label{eq:app-D}
	D = \frac{1}{2}
\end{equation}
holds for the diffusion coefficient.

In the Fokker--Planck picture, the motion of a singular `point'
described by eq.~(\ref{eq:app-y-equation}) can be recast into the form
of a flow of the distribution of such points.
Let us introduce the distribution function $f(y)$
that correspond to uniformly distributed
coordinate $x$ in the range $0 \leq x \leq L$, i.e.,
$f(y){\rm d}y$ is the probability that $y(x)$ falls into the range
$[y,y+{\rm d}y]$ when $x$ is distributed uniformly,
\begin{equation}
  f(y){\rm d}y = \frac{1}{L} {\rm d}x \; .
  \label{eq:app-f-dy}
\end{equation}
Since the function $f(y)$ describes the `time'-averaged distribution,
it must remain unchanged under the motion governed
by eq.~(\ref{eq:app-y-equation}).
Therefore, the continuity equation dictates that
\begin{equation}
  \frac{{\rm d}j(y)}{{\rm d}y} = 0 \; ,
  \label{eq:app-dj-dy}
\end{equation}
where $j(y)$ is the flow of points due to drift and diffusion,
\begin{equation}
  j(y) = v(y)f(y) - D\frac{{\rm d}f(y)}{{\rm d}y} \; .
  \label{eq:app-j}
\end{equation}
Substituting the drift velocity $v(y)$ and the diffusion coefficient~$D$
form eqs.~(\ref{eq:app-v-def}) and~(\ref{eq:app-D}),
one obtains a differential equation for distribution
function $f(y)$~\cite{vanKampen},
\begin{equation}
  (y^2+e) f(y) - \frac{1}{2} \frac{{\rm d}f(y)}{{\rm d}y} = c \; ,
  \label{eq:app-f-equation}
\end{equation}
where $c$ is a constant that is determined
from the boundary conditions below.

To find the unique solution $f(y)$ of this equation
we employ the boundary conditions at $y\to\pm\infty$.
It is evident from eq.~(\ref{eq:app-y-def}) that large $|y|$
correspond to small $\psi(x)$, i.e., to the neighborhoods of the
zeros $x_0,\ldots,x_n$ of the wave function $\psi(x)$.
Keeping only the linear term in the Taylor series around $r$th zero, $x_r$,
we obtain
\begin{equation}
  y(x) = -\frac{\psi'(x)}{\psi(x)} \approx
  -\frac{\psi'(x_r)}{(x-x_r)\,\psi'(x_r)} = \frac{1}{x_r-x} \; .
  \label{eq:app-y-around-zero}
\end{equation}
Using the equations~(\ref{eq:app-f-dy}) and~(\ref{eq:app-y-around-zero}),
we see that $f(y) \approx 1/(Ly^2)$ for large $y^2$.
This result must be multiplied by the number of zeros $n$
because each of them gives a contribution to $f(y)$.
We recall that $n(e) =L N(e)$ from eq.~(\ref{eq:app-N-def})
and obtain the boundary conditions
\begin{equation}
  f(y) \to \frac{N(e)}{y^2}  \; \text{for} \; |y| \to \infty \; .
  \label{eq:app-y-BC}
\end{equation}
Inserting this expression into eq.~(\ref{eq:app-f-equation})
leads to the value of the constant $c$,
\begin{equation}
  c = N(e) \; .
   \label{eq:app-c}
\end{equation}
It is worth noting that the `flow' $c$ has a meaning of the
inverse average distance between zeros of the wave functions, i.e.,
the inverse average `time' of the motion from $y=-\infty$ to $y=+\infty$.

The first-order differential equation~(\ref{eq:app-f-equation})
with $c = N(e)$ and boundary conditions~(\ref{eq:app-y-BC})
is readily solved by the method of undetermined coefficients,
\begin{eqnarray}
  f(y) &=& 2N(e) \exp\left(\frac{2y^3}{3}+2e y\right)
  \nonumber   \\
  &&	\times \int_y^{+\infty} \exp\left(-\frac{2u^3}{3}-2e u\right)
              {\rm d}u \; .
        \label{eq:app-f}
\end{eqnarray}
Finally, $N(e)$ can be calculated by
inserting~(\ref{eq:app-f}) into the normalization condition
for the distribution $f(y)$,
\begin{equation}
  \int_{-\infty}^{+\infty} f(y) {\rm d}y = 1 \; .
  \label{eq:app-f-normalization}
\end{equation}
To evaluate eq.~(\ref{eq:app-f-normalization}),
we must perform the integrations over $u$ and $y$.
It is convenient to make the substitution $u = y + t$
in eq.~(\ref{eq:app-f}) after which the integral over the new variable $t$
acquires constant limits, $0 \leq t <\infty$.
Switching the order of integrations gives rise to a Gaussian integral
over $y$. After integration over~$y$,
the normalization condition~(\ref{eq:app-f-normalization}) takes the form
\begin{equation}
  \sqrt{2\pi} N(e) \int_0^{\infty}
  \frac{{\rm d}t}{\sqrt{t}} \exp\left( - \frac{t^3}{6} - 2e t \right)
  = 1 \; .
  \label{eq:app-f-normalization-2}
\end{equation}
The remaining integral can be expressed using Airy functions $\Ai(x)$
and $\Bi(x)$ of the
first and the second kind~\cite[p.~32]{Muldoon1977}.
This leads to expression~(\ref{eq:IntegratedexactDOSasymptotic})
for the integrated particle density $N(e)$.

The result~(\ref{eq:IntegratedexactDOSasymptotic}) is valid
for a certain set of unit lengths and energies
in which the Schr\"odinger equation has the dimensionless form
of eq.~(\ref{eq:app-SE}), and the statistics of the potential
is also dimensionless, see eq.~(\ref{eq:app-V-statistics}).
However, it is easy
to rewrite eq.~(\ref{eq:IntegratedexactDOSasymptotic}) in arbitrary units.
To do this, it is enough to insert a factor of dimension (energy)$^{-1}$
into the arguments of the Airy functions,
and to multiply the whole expression by a factor of
dimension (length)$^{-1}$. These dimension factors
must be products of appropriate powers of the
quantities $\hbar$, $2m$, and $S$,
where $S$ is defined by eq.~(\ref{eq:white_noise}).
The resulting expression reads
\begin{eqnarray}
  N(E)&=& \frac{1}{\pi^2} \sqrt[3]{\frac{2Sm^2}{\hbar^4}}
  \Biggl\{
  \left[\Ai\left(-\sqrt[3]{\frac{2\hbar^2}{mS^2}} \; E\right)\right]^2
  \nonumber \\
&&  + \left[\Bi\left(-\sqrt[3]{\frac{2\hbar^2}{mS^2}} \; E\right)\right]^2
  \Biggr\}^{-1}.
  \label{eq:app-IntegratedexactDOSasymptotic}
\end{eqnarray}
Here, all quantities are expressed in SI units.

\subsection{Localization length}

In one dimension, all electron eigenstates are Anderson-localized. A wave function $\psi(x)$ of such an eigenstate has exponential tails at $x \to \pm\infty$:
\begin{equation}
\psi(x) \simeq \exp(-|x|/a),
\end{equation}
where coordinate $x$ is counted from the center of the eigenstate, and $a$ is the localization length.

Here we calculate $a$ as a function of the electron energy in a random white-noise potential.

We consider the left tail ($x<0$), where $\ln \psi(x) \approx (x/a) + const$. The quantity $1/a$ is therefore equal to
\begin{equation}
\frac1a = \lim_{x\to-\infty} \frac{\ln[\psi(0)] - \ln[\psi(x)]}{0 - x}  = \left\langle \frac{\mathrm d \ln\psi(x)}{\mathrm d x} \right\rangle_{x\in(-\infty,0)} .
\end{equation}
%Following Appendix A of Paper 1, we introduce function
%\begin{equation}
%y(x) = - \frac{\psi'(x)}{\psi(x)} \equiv - \frac{\mathrm d \ln\psi(x)}{\mathrm d x} \,,
%\end{equation}
In terms of the function $y(x)$ determined by eq.~(\ref{eq:app-y-def}),
\begin{equation} \label{eq:a_via_f}
\frac1a = - \langle y \rangle = - \int_{-\infty}^{+\infty} y \, f(y) \, \mathrm d y ,
\end{equation}
where $f(y)$ is the distribution function of $y$ determined by eq.~(\ref{eq:app-f}). Substituting this expression into eq.~(\ref{eq:a_via_f}), one can calculate the localization length $a$. A convenient way for this calculation is to make the substitution $u = y + t$, then integrate over $y$ from $-\infty$ to $+\infty$, and finally integrate over $t$ from 0 to $+\infty$. The result reads:
\begin{equation} \label{eq:a_answer1}
\frac1a = - \frac{\pi^2}{2^{5/3}} \, N(e) \, \frac{\mathrm d \left[   [\Ai(-2^{2/3}e)]^2+ [\Bi(-2^{2/3}e)]^2   \right]}{\mathrm d e} \,,
\end{equation}
where $e$ is the dimensionless energy, and $N(e)$ is the integrated DOS. The combination of Airy functions $\Ai$ and $\Bi$ in the latter equation is related to the integrated DOS by eq.~(\ref{eq:IntegratedexactDOSasymptotic}):
\begin{equation} \label{eq:Airy_via_N}
[\Ai(-2^{2/3}e)]^2+ [\Bi(-2^{2/3}e)]^2 = \frac{1}{2^{1/3} \pi^2 N(e)} \,.
\end{equation}
Combining equations~(\ref{eq:a_answer1}) and~(\ref{eq:Airy_via_N}), one can get the (dimensionless) localization length $a$ as a function of the dimensionless energy $e$:
\begin{equation} \label{eq:a_via_e}
a(e) = 4 \left[ \frac{\mathrm d \ln N(e)}{\mathrm d e} \right]^{-1} ,
\end{equation}
where $N(e)$ is given by eq.~(\ref{eq:IntegratedexactDOSasymptotic}).

One can rewrite this expression in physical (SI) units by replacing $\mathrm d/\mathrm d e$ with ${\cal T}(\mathrm d/\mathrm d E)$ and multiplying the right-hand side by $\ell_{\rm wp}$, where $\cal T$ and $\ell_{\rm wp}$ are given by eq.~(\ref{eq:app-1d-values}), and $E$ is the energy in physical units:
\begin{equation} \label{eq:a_via_E}
a(E) = \frac{2\hbar^2}{mS} \left[ \frac{\mathrm d \ln N(E)}{\mathrm d E} \right]^{-1} .
\end{equation}
Function $N(E)$ here is given by eq.~(\ref{eq:app-IntegratedexactDOSasymptotic}).

Let us consider the limiting cases. At high energies, $E \to +\infty$, the integrated DOS $N(E)$ is proportional to $\sqrt E$, just as in a one-dimensional conduction band in the absence of disorder. Hence, $\mathrm d \ln N(E)/\mathrm d E = 1/2E$, and
\begin{equation} \label{eq:a_via_E_high_energy}
a(E) = \frac{4\hbar^2E}{mS} \, .
\end{equation}

At low energies $E \to -\infty$, i.~e. deeply in the band tail, one can omit $\Ai$ in eq.~(\ref{eq:app-IntegratedexactDOSasymptotic}), and use asymptotic formula $\Bi(x) \approx \exp(\frac23 x^{3/2}) / (\sqrt\pi x^{1/4})$, which gives $N(E) \sim \exp\left[ \frac43 \sqrt{\frac{2\hbar^2}{mS^2}} (-E)^{3/2} \right]$. Substitution of this estimate into eq.~(\ref{eq:a_via_E}) yields
\begin{equation} \label{eq:a_via_E_low_energy}
a(E) = \frac{\hbar}{\sqrt{-2mE}} = \frac{1}{k} \, .
\end{equation}

It is well-known~\cite{Thouless1972} that the localization length $a(E)$ in 1D can be expressed via the integrated density of states $N(E)$ in an integral form. It follows from the results of Ref.~\cite{Thouless1972} that
\begin{equation} \label{eq:a_via_E_Thouless}
\frac{1}{a(E)} = \int_{-\infty}^{+\infty} \frac{N(E')-N_0(E')}{E-E'} \, \mathrm d E' + \kappa_0(E) ,
\end{equation}
where the Cauchy principal value of the integral is assumed, $N_0(E)$ is the free-particle integrated density of states,
\begin{equation} \label{eq:free-particle-DOS}
N_0(E) =
\begin{cases}
\;\;\;\; 0 & \text{ if } E \le 0 , \\
\frac{\sqrt{2mE}}{\pi\hbar}  & \text{ if } E \ge 0 ,
\end{cases}
\end{equation}
and $\kappa_0(E)$ is the inverse localization length of a particle in zero barrier potential,
\begin{equation} \label{eq:kappa0}
\kappa_0(E) =
\begin{cases}
\frac{\sqrt{-2mE}}{\hbar} & \text{ if } E \le 0 , \\
\;\;\;\;\; 0  & \text{ if } E \ge 0 .
\end{cases}
\end{equation}
We have checked numerically that both expressions for localization length $a(E)$, eq.~(\ref{eq:a_via_E}) and eq.~(\ref{eq:a_via_E_Thouless}), yield identical results for the white-noise integrated density of states $N(E)$ given by eq.~(\ref{eq:app-IntegratedexactDOSasymptotic}).

The advantage of our eq.~(\ref{eq:a_via_E}) is that it provides an expression for the localization length in a closed form, rather than in a form of an integral. On the other hand, this expression is valid only in the case of a white-noise random potential. For other 1D potentials, the more complicated eq.~(\ref{eq:a_via_E_Thouless}) should be used.

\section{Energy and length scales}
\label{app:B}

In Section~\ref{subsec:Model}, we considered features
of a particle in a random potential such as the typical width
of the wave packet~$\ell_{\rm wp}$, the typical kinetic energy
of a localized state~$\cal T$, and the strength of the potential~$S$.
The latter two quantities are defined in eqs.~(\ref{eq:estimatecalT-or-ellwp})
and~(\ref{eq:white_noise}), respectively. Then, we introduced a dimensionless
representation, in which $\ell_{\rm wp}$ and $\cal T$
are used as units of length and energy.
The aim of this appendix is to fix the values of $\ell_{\rm wp}$ and $\cal T$
so that the strength of the potential becomes equal to unity in dimensionless units.

To achieve this goal, we present a dimensionless version
of eq.~(\ref{eq:white_noise})
that defines the dimensionless disorder strength $S_L$,
\begin{equation}
 \langle \bar{\cal V}_{\rm R}(\vec{x}\,) \bar{\cal V}_{\rm R}(\vec{x}\,{}^{\prime})
 \rangle_{\rm R} =
 S_L \delta\left(\vec{x}-\vec{x}\,{}^{\prime}\right)\; ,
 \label{eq:app-SL-def}
\end{equation}
where $\vec{x} = \vec{r}/\ell_{\rm wp}$ and $\bar{\cal V}_{\rm R}
= {\cal V}_{\rm R} / {\cal T}$ are scaled coordinates and potentials.
We divide both sides of eq.~(\ref{eq:app-SL-def}) by the corresponding sides
of eq.~(\ref{eq:white_noise}), and take into account that
$\delta\left(\vec{x}-\vec{x}\,{}^{\prime}\right)
= \delta\left[ (\vec{r}-\vec{r}\,{}^{\prime})/\ell_{\rm wp} \right] =
\ell_{\rm wp}^d \delta\left(\vec{r}-\vec{r}\,{}^{\prime}\right)$ and obtain
\begin{equation}
  \frac{1}{{\cal T}^2} = \frac{S_L}{S} \, \ell_{\rm wp}^d
  \label{eq:app-proportion}
\end{equation}
in $d$ dimensions. Substituting~$\cal T$ from eq.~(\ref{eq:estimatecalT-or-ellwp}),
we express the dimensionless strength of the disorder potential as
\begin{equation}
  S_L = \frac{S}{{\cal T}^2 \ell_{\rm wp}^d} = S  \frac{(2m)^2
    \ell_{\rm wp}^{4-d}}{\hbar^4} \; .
  \label{eq:app-SL-value}
\end{equation}
Now we assume $S_L = 1$ which yields
\begin{equation}
  \ell_{\rm wp} = \left( \frac{\hbar^4}{4m^2S} \right)^{1/(4-d)}
  \label{eq:app-ell-wp-value}
\end{equation}
for the width of the wave packet and
\begin{equation}
  {\cal T} = \frac{\hbar^2}{2m\ell_{\rm wp}^2} =
  \left( \frac{2m}{\hbar^2} \right)^{d/(4-d)}  S^{2/(4-d)}
  \label{eq:app-T-value}
\end{equation}
for the typical kinetic energy of a localized state.
In one dimension ($d=1$),
\begin{equation}
  \ell_{\rm wp} = \frac{\hbar^{4/3}}{(2m)^{2/3} S^{1/3}} \quad ,  \quad
  {\cal T} = \frac{(2m)^{1/3} S^{2/3}}{\hbar^{2/3}} \; .
  \label{eq:app-1d-values}
\end{equation}
Note that the expressions~(\ref{eq:app-ell-wp-value})--(\ref{eq:app-1d-values})
are unique combinations of $[2m] = \text{M}$, $[\hbar] = \text{M}\text{L}^2/\text{T}$
and $[S] = \text{E}^2\text{L}^d = \text{M}^2 \text{L}^{4+d}/\text{T}^4$
that possess the correct dimensions.
%in $\text{M}=\text{kg}$, $\text{L}=\text{m}$, and$\text{T}=\text{s}$.

%\newpage
% \bibliographystyle{unsrtnat}
\bibliography{LL_filter}

\end{document}